\PassOptionsToPackage{normalem}{ulem}
\documentclass[]{arxiv_template}  
\usepackage{enumitem}
\usepackage[toc,page,header]{appendix}
\usepackage[utf8]{inputenc}
\usepackage[T1]{fontenc}
\usepackage{hyperref}
\usepackage{url}
\usepackage{booktabs}
\usepackage{amsfonts}
\usepackage{nicefrac}
\usepackage{float}
\usepackage{textcomp}
\definecolor{orange-web}{RGB}{255, 165, 0}
\definecolor{sagegreen}{RGB}{120, 150, 120}
\usepackage{makecell}
\usepackage{microtype}
\usepackage{graphicx}
\usepackage{breqn}
\usepackage{algorithm}
\usepackage{algorithmic}
\usepackage{tabularx}
\usepackage{multirow}
\usepackage{pifont}
\usepackage{amsmath,amssymb,amsthm}
\usepackage{etoolbox}
\usepackage{lipsum}  
\usepackage{minitoc}
\usepackage{tablefootnote}
\usepackage{threeparttable}
\usepackage{wrapfig}
\usepackage{appendix}
\usepackage{multirow}
\usepackage{colortbl}
\usepackage{xspace}
\usepackage{longtable}
\usepackage{color}
\usepackage{xcolor}
\usepackage[most]{tcolorbox}
\usepackage[utf8]{inputenc}
\usepackage[most]{tcolorbox}
\usepackage{listings}
\usepackage{fontawesome5}  
\usepackage{xcolor}       % 用于定义颜色
\usepackage{caption}
\usepackage{listings}
\definecolor{boxback}{gray}{0.95}
\definecolor{promptback}{gray}{0.98}
\definecolor{outputback}{RGB}{230, 245, 255}
\definecolor{titleblue}{RGB}{0, 110, 180}

\newcommand{\reduction}[1]{\textcolor{green!60!black}{\small $\downarrow$#1}}
\newcommand{\increase}[1]{\textcolor{green!60!black}{\small $\uparrow$#1}}
\newcommand{\decrease}[1]{\textcolor{red!60!black}{\small $\downarrow$#1}}

\newcommand{\approach}{{SWE-Pruner}\xspace}

\newtcolorbox{casestep}[2][]{
    enhanced,
    breakable,
    colback=boxback,
    colframe=gray!70,
    coltitle=white,
    colbacktitle=titleblue,
    fonttitle=\bfseries,
    title=#2,
    attach boxed title to top left={yshift=-2mm, xshift=4mm},
    boxed title style={sharp corners, boxrule=0pt},
    before skip=15pt,
    #1
}
\lstdefinestyle{pythoncode}{
    language=Python,
    basicstyle=\large\ttfamily,
    keywordstyle=\color{blue!60!cyan!90}\bfseries,
    commentstyle=\color{gray!60}\itshape,
    stringstyle=\color{teal!70!green!80},
    identifierstyle=\color{violet!80!black},
    backgroundcolor=\color{blue!3!white},
    frame=single,
    frameround=tttt,
    framesep=4pt,
    rulecolor=\color{blue!20!gray!30},
    numbers=none,
    breaklines=true,
    showstringspaces=false,
    tabsize=2,
    captionpos=b,
    xleftmargin=8pt,
    xrightmargin=8pt
}
\usepackage{amsmath}
\allowdisplaybreaks  % 允许 align 等环境分页
\lstdefinestyle{promptstyle}{
    backgroundcolor=\color{promptback},
    basicstyle=\ttfamily\small,
    breaklines=true,
    frame=tb,
    framerule=0pt,
    rulecolor=\color{gray!40},
    xleftmargin=1em,
    xrightmargin=1em,
      xleftmargin=0pt,            % 左边无额外缩进
      xrightmargin=0pt,           % 右边无额外缩进
      framesep=3pt,               % 框线与文字间距，可自己调
      rulesep=2pt,
}

\newtcolorbox{modeloutput}[2][]{
    enhanced,
    breakable,
    colback=outputback,
    colframe=titleblue!50,
    boxrule=1pt,
    fonttitle=\bfseries\small,
    coltitle=titleblue,
    title=#2,
    #1
}
\newtcolorbox{modeloutput_continue}[1][]{
    enhanced,
    breakable,
    colback=outputback,
    colframe=titleblue!50,
    boxrule=1pt,
    #1
}
\title{SWE-Pruner: Self-Adaptive Context Pruning for Coding Agents} 

\makeatletter
\gdef\authorlist{%
  \begin{center}
    \authorfont\sffamily
    Yuhang Wang$^{1*}$,  
    Yuling Shi$^{1*}$,
    Mo Yang$^{2}$,
    Rongrui Zhang$^{1}$,
    Shilin He$^{3}$\\
    Heng Lian$^{1}$, 
    Yuting Chen$^{1}$, 
    Siyu Ye$^{3}$, 
    Kai Cai$^{3}$,
    Xiaodong Gu$^{1\dag}$
  \end{center}
}
\makeatother

\makeatletter
\gdef\affiliationlist{%
  \begin{center}
    \affiliationfont\sffamily
    $^{1}$LLMSE Lab, Shanghai Jiao Tong University
    $^{2}$Sun Yat-sen University
    $^{3}$Douyin Group
  \end{center}
}
\makeatother

\abstract{
    LLM agents have demonstrated remarkable capabilities in software development, but their performance is hampered by long interaction contexts, which incur high API costs and latency. 
While various context compression approaches such as LongLLMLingua have emerged to tackle this challenge, they typically rely on fixed metrics such as PPL, ignoring the task-specific nature of code understanding. As a result, they frequently disrupt syntactic and logical structure and fail to retain critical implementation details.
In this paper, we propose \approach, a self-adaptive context pruning framework tailored for coding agents. Drawing inspiration from how human programmers ``selectively skim'' source code during development and debugging, \approach performs task-aware adaptive pruning for long contexts. Given the current task, the agent formulates an explicit goal (e.g., ``focus on error handling'') as a hint to guide the pruning targets. A lightweight \textit{neural skimmer} (0.6B parameters) is trained to dynamically select relevant lines from the surrounding context given the goal. 
Evaluations across four benchmarks and multiple models validate \approach's effectiveness in various scenarios, achieving 23-54\% token reduction on agent tasks like SWE-Bench Verified while even improving success rates, and up to 14.84$\times$ compression on single-turn tasks like LongCodeQA with minimal performance impact.
}

\titleformat*{\paragraph}{\bfseries}
\checkdata[Project Page]{\url{https://github.com/Ayanami1314/swe-pruner}}

\begin{document}
\maketitle
\tcbset{reset}

{
\renewcommand{\thefootnote}{\fnsymbol{footnote}}
  \footnotetext[1]{Equal Contribution} 
  \footnotetext[2]{Corresponding author: xiaodong.gu@sjtu.edu.cn}
}

\begin{figure}[H]
  \centering
  \subfloat[Claude Sonnet 4.5]{\includegraphics[width=0.48\textwidth]{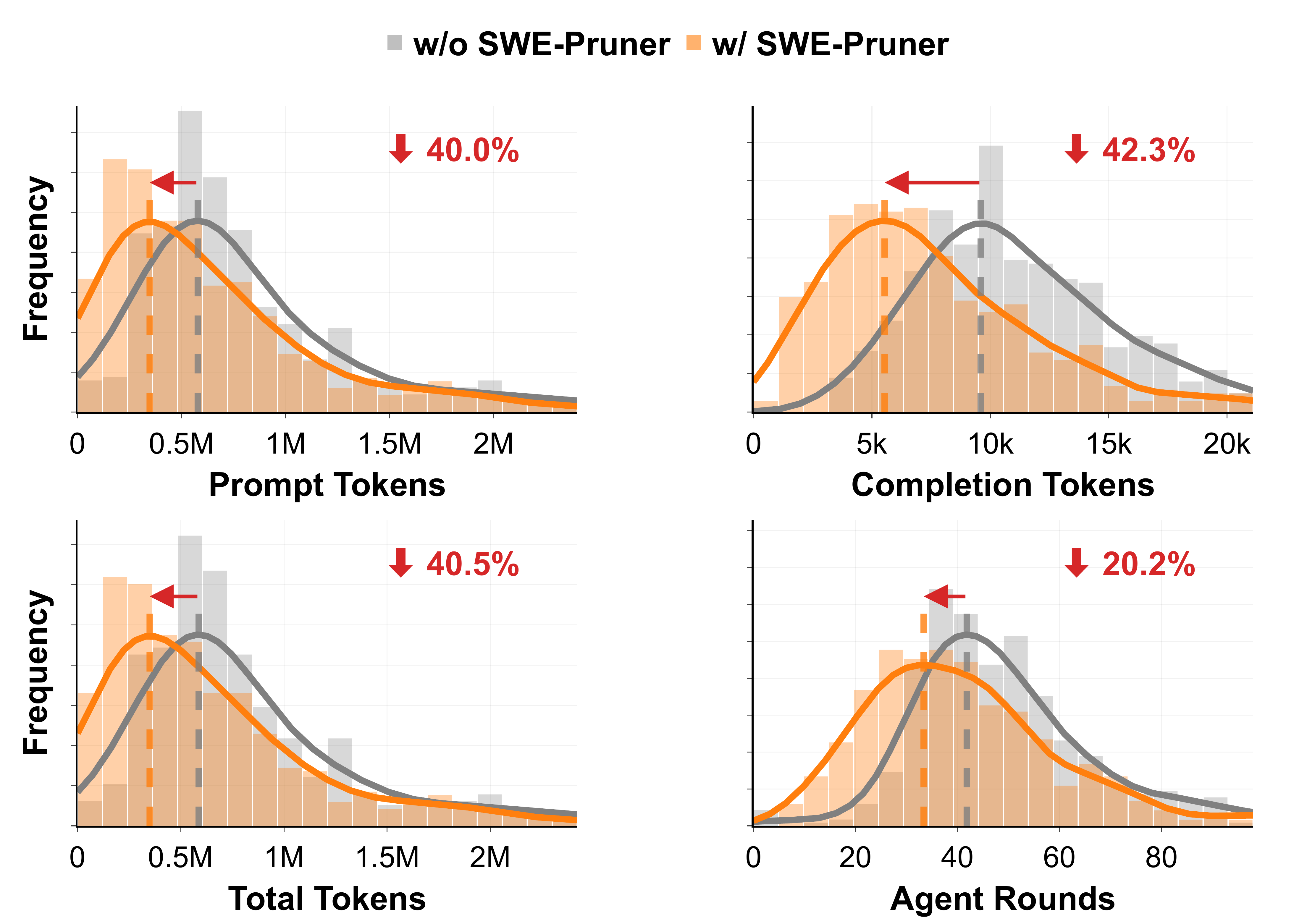}\label{fig:claude_compare_full}}
  \subfloat[GLM 4.6]{\includegraphics[width=0.48\textwidth]{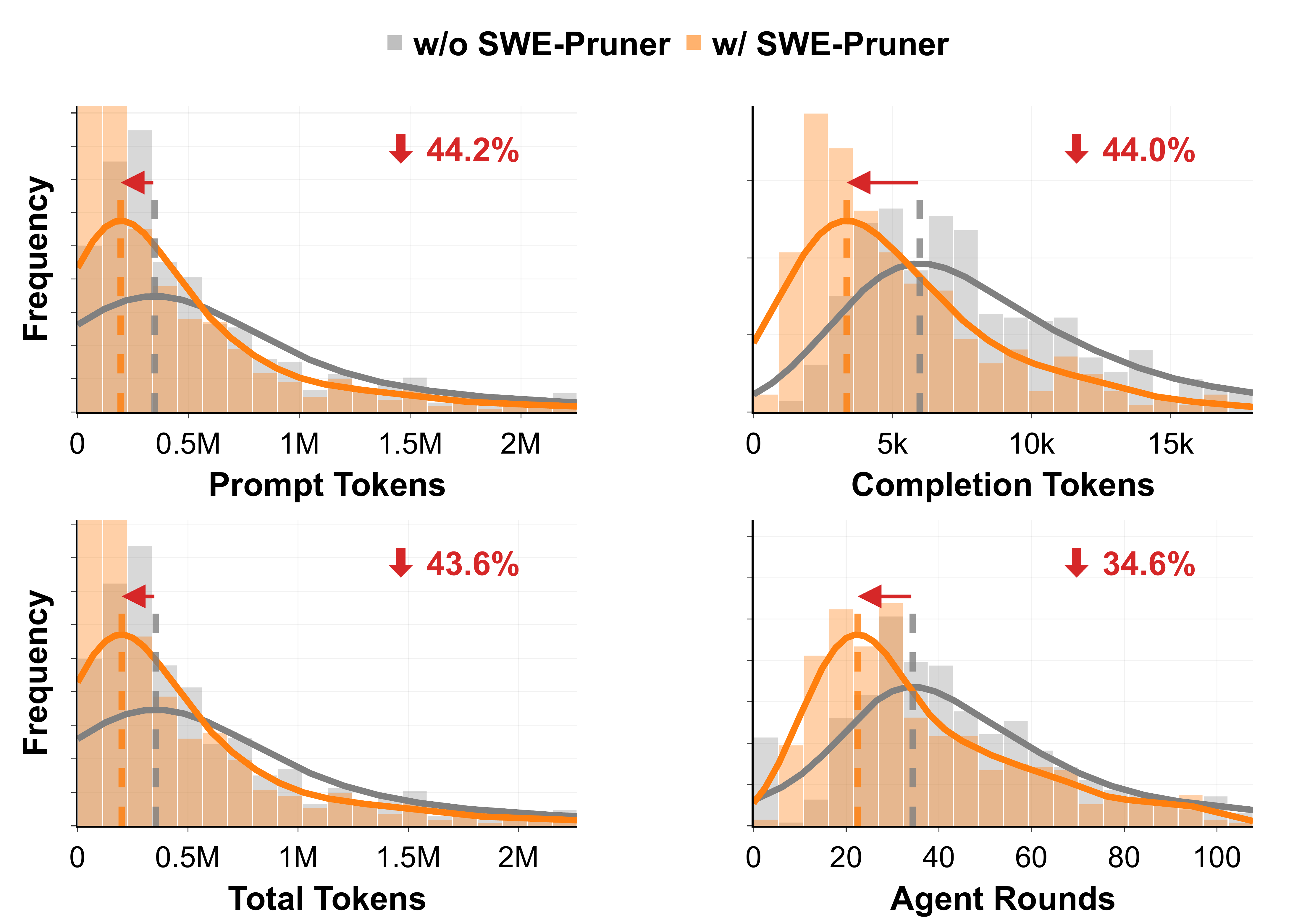}\label{fig:glm_compare_full}}
  \hfill
  \caption{\textbf{Efficiency analysis on SWE-Bench Verified.} \approach (\textcolor{orange}{orange}) achieves substantial reductions on Token Cost and Agent Rounds for the base Mini SWE Agent (\textcolor{gray}{gray}) for both Claude Sonnet 4.5 and GLM 4.6.}
  % (a) With Claude Sonnet 4.5: 38.7\% in prompt tokens, 40.8\% in completion tokens, 39.2\% in total tokens, and 18.3\% in agent rounds. (b) With GLM 4.6: 44.2\% in prompt tokens, 44.0\% in completion tokens, 43.6\% in total tokens, and 34.6\% in agent rounds. }
  \label{fig:claude_token_count}
\end{figure}

% \begin{figure}[H]
%   \centering
%   \includegraphics[width=0.45\textwidth]{figures/comparisor.png}
%   \vspace{-0.26cm}
%   \caption{\textbf{\approach effectiveness on SWE-Bench Verified with 
%   Claude Sonnet 4.5.} Our \approach (\textcolor{orange}{orange}) saves 39.2\% in total tokens and 18.3\% in agent rounds for the original Mini SWE Agent (\textcolor{blue}{blue}), demonstrating substantial efficiency gains while maintaining task performance.}
%   \label{fig:claude_token_count}
% \end{figure}

\section{Introduction}
\label{sec:intro}

Large Language Models (LLMs) have rapidly progressed in software engineering tasks, from simple code understanding~\citep{chen2021evaluating,li2023starcoder,yang2025elaboration,hu2025flowmaltrans,zeng2026codesumm} to interactive agents capable of navigating repositories, running tests, and submitting patches end-to-end~\citep{shi2024code,chen2025swe,li2025swe,xiao2025improving,wang2026effiskill}. Recent coding agents like Claude Code~\citep{claude_code} and Gemini CLI~\citep{gemini_cli} have augmented LLMs with sophisticated toolchains including terminals, editors, and file search, enabling multi-step reasoning over complex codebases.

Despite this progress, a critical bottleneck remains: the accumulation of context length within the constrained window of LLMs. Navigating real-world software repositories confronts agents with a massive ``Context Wall.'' Although long-context models have emerged~\citep{achiam2023gpt,team2024gemini}, blindly ingesting large volumes of code incurs prohibitive inference costs~\citep{jiang2023llmlingua} and introduces severe noise, leading to attention dilution and hallucinations~\citep{liu2023lost,li2023loogle,wang2025position,wang2026fasa}. 

Although context compression techniques offer a potential remedy, existing approaches face critical limitations when applied to coding agents. Prior work on context compression primarily targets natural language~\citep{jiang2023llmlingua,li2023compressing,jiang2024longllmlingua,mu2023learning}, leading to severe trade-offs when applied to code~\citep{shi2025longcodezip}: token pruning methods compromise syntactic validity, while abstractive techniques discard character-level information critical for debugging~\citep{jiang2023llmlingua,li2023compressing}. Beyond structural concerns, these methods are fundamentally misaligned with coding agent requirements—they operate with static compression ratios and task-agnostic criteria (e.g., perplexity), unable to dynamically prioritize context based on the evolving goals of multi-turn agent interactions~\citep{zhang2024hierarchical,yang2024less}. 

% Although context compression techniques offer a potential remedy, most existing approaches (e.g., LLMLingua and Selective Context) are predominantly tailored for natural language inputs~\citep{jiang2023llmlingua,li2023compressing,jiang2024longllmlingua,mu2023learning}. Adapting them to coding agent introduces significant challenges. First\shi{fix this later}, existing methods typically rely on a fixed compression criteria, such as perplexity (PPL), which fail to account for the dynamic context requirements inherent in specific coding tasks~\citep{jiang2023llmlingua,li2023compressing}. As such, they face inherent trade-offs when applied to code: token-level compression methods often disrupt syntax, leading to invalid code fragments, while summarization-based methods often omit character-level details that are critical for debugging and program analysis~\citep{zhang2024hierarchical,yang2024less}. 

To address these limitations, we introduce \approach, a self-adaptive pruning framework tailored for coding agents. Our key insight stems from how both programmers and agents navigate large codebases: rather than reading code line by line, they employ \emph{goal-driven selective attention} that programmers quickly scan to locate relevant sections based on their goals (e.g., ``find error handling logic''), while agents use tools like \texttt{grep} or file search to coarsely retrieve targeted code regions including excessive surrounding context. \approach refines this process by enabling agents to articulate explicit natural language goals alongside retrieval actions. To capture the dynamic, task-specific objectives of agent workflows, we train a lightweight 0.6B pruning model on 61K synthetic data, enabling adaptive selection conditioned on given goals. By operating at line-level granularity, our approach preserves syntactic and structural integrity while achieving precise compression that distills coarse retrieval results into focused, task-relevant context.

We evaluate \approach across multi-turn agent tasks (SWE-Bench Verified~\citep{jimenez2024swe} and SWE-QA~\citep{peng2025swe}) and single-turn long-context understanding tasks (Long Code Completion~\citep{guo2023longcoder} and Long Code QA~\citep{rando2025longcodebench}). Across models and benchmarks, \approach consistently delivers substantial efficiency gains while maintaining or even improving task performance. Beyond token savings (e.g., 39\% reduction on SWE-Bench Verified with Claude Sonnet 4.5, as illustrated in \Cref{fig:claude_token_count}), the focused context also improves agent decision quality, reducing interaction rounds by up to 26\% through more decisive reasoning and fewer redundant exploratory actions.

In summary, we make the following contributions:
\begin{itemize}
    \item We propose a novel pruning framework for coding agents that performs task-aware, line-level pruning to alleviate the context wall problem.
    % Our framework is universally applicable to both agentic (multi-turn) and non-agentic (single-turn) coding tasks.
    \item We design a novel architecture with a lightweight backbone of only 0.6B parameters for efficient inference with adaptive, task-dependent pruning.
    % \item Extensive experiments demonstrate that \approach achieves substantial efficiency gains with minimal performance degradation.
    \item Evaluations across various benchmarks demonstrate 23--54\% token reduction on agent tasks and up to 14.8$\times$ compression on single-turn tasks with minimal performance impact.
\end{itemize}

\section{Motivation}
% @ chenfei

\begin{figure}[t]
    \centering
    \includegraphics[width=0.45\linewidth]{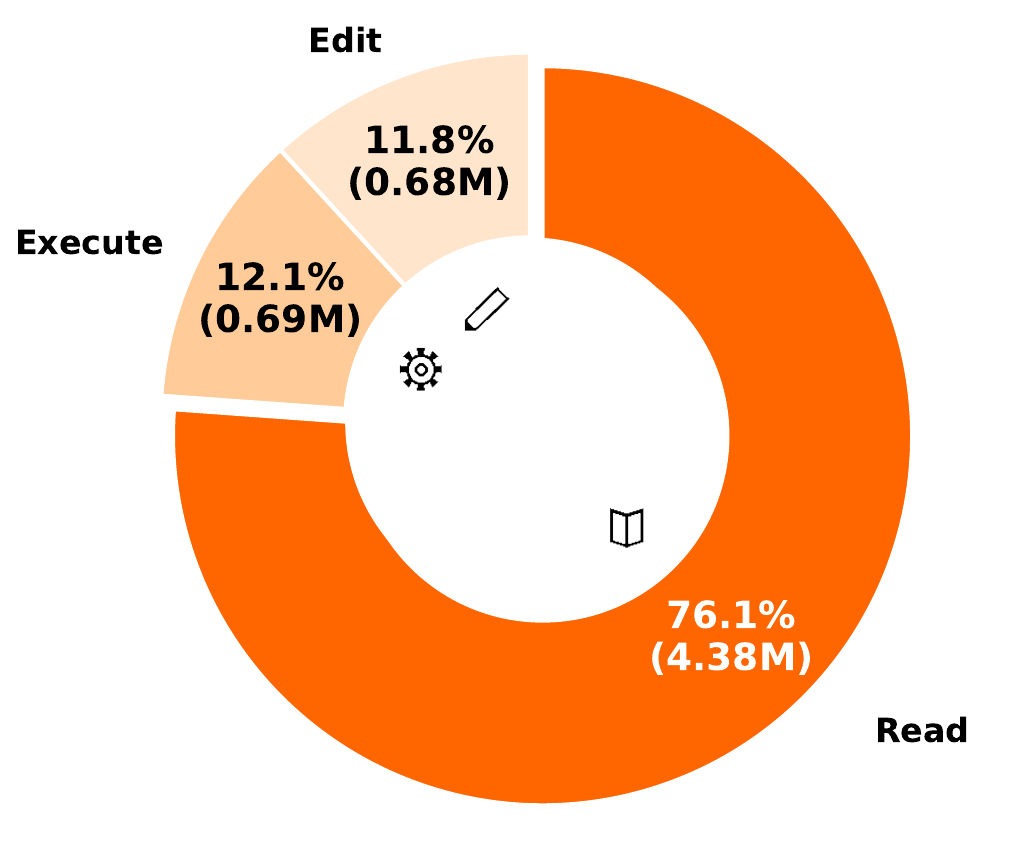}
    \caption{Token cost distribution over different tool calls for Mini-SWE-Agent on SWE-Bench Verified with Claude Sonnet 4.5. Read operations dominate token consumption at 76.1\%, motivating the need for context pruning mechanisms.}
    \vspace{-0.25cm}
    \label{fig:agent_token_count}
  \end{figure}

  \begin{figure*}[t]
    \centering
    \includegraphics[width=0.85\linewidth]{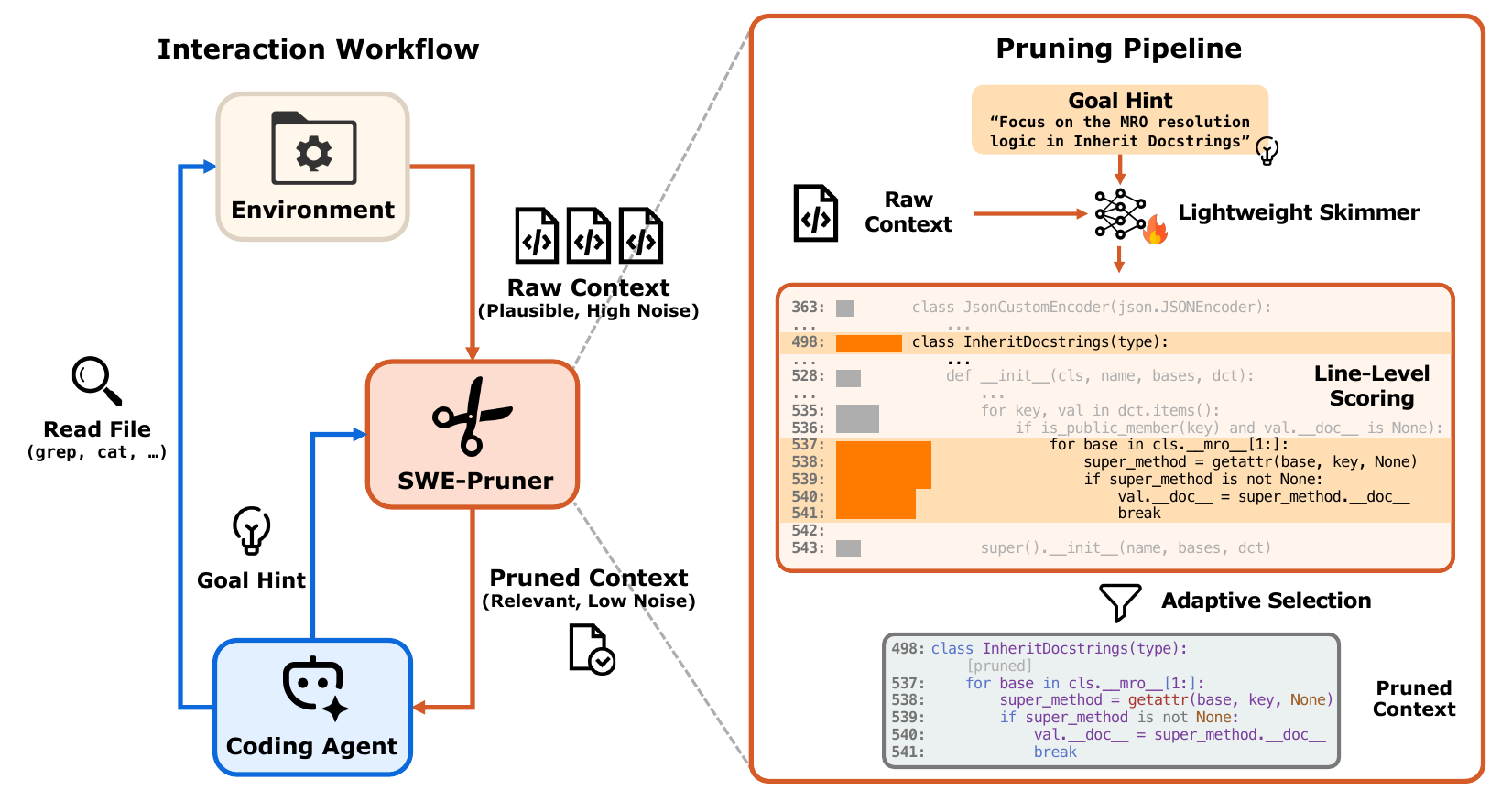}
    \caption{Overview of \approach. Left: The \textbf{Interaction Workflow} demonstrates how \approach functions as a middleware between the Coding Agent and the Environment. It intercepts the Raw Context from file operations and delivers a Pruned Context to the agent. Right: The \textbf{Pruning Pipeline} details the internal mechanism. Based on a specific goal hint from the coding agent, the \emph{neural skimmer} processes the raw context through line-level scoring and adaptive selection to deliver the pruned context.}
    \label{fig:overview}
    \vspace{-0.25cm}
  \end{figure*}
  
  In practice, coding agents spend an excessive amount of their token budget on repeatedly exploring the codebase~\citep{majgaonkar2025understanding,xiao2025improving}.
  To quantify this, we perform a preliminary analysis of the trajectories by a Mini SWE Agent~\citep{yang2024sweagent}. The agent is instantiated with Claude Sonnet 4.5 on SWE-Bench Verified. We extract action types from the trajectories and categorize them into three types: 1) \emph{read}, which denotes file and directory inspection operations such as \texttt{cat}, \texttt{grep}, and \texttt{head}; 2) \emph{execute}, which means running programs or scripts for testing, and 3) \emph{edit}, representing in-place modifications. We count the total tokens consumed in each agent round. As shown in \Cref{fig:agent_token_count}, \emph{read}-type operations consume an overwhelming 76.1\% of total tokens, significantly exceeding \emph{execute} (12.1\%) and \emph{edit} (11.8\%) operations combined. This issue is further exacerbated in multi-round agent interactions, where code retrieved in earlier rounds persists in the context and accumulates as the interaction progresses. Similar patterns are observed with other backbone models like GLM-4.6 (\Cref{app:glm_analysis}).
  
  This empirical pattern reveals a critical opportunity for context optimization. When navigating unfamiliar codebases, agents must extensively explore through coarse-grained file operations that stream entire files or blocks into the context. While this exploratory strategy is necessary for understanding the codebase structure, it causes substantial token costs and introduces redundant content that accumulates across rounds. Addressing this challenge requires a context pruning mechanism that is context-aware (capable of identifying relevant information based on the agent's current goal) and lightweight enough to avoid introducing additional latency. These observations strongly motivate \approach, an lightweight pruning framework that enables more efficient token allocation throughout the problem-solving trajectory.
  
  \section{Approach}
  \label{sec:approach}
  
  \subsection{Overview}
  We introduce \approach, a framework that prunes long code contexts for agents through task-aware, adaptive filtering. As illustrated in \Cref{fig:overview}, \approach operates as middleware between coding agents and their environment. When an agent issues file-reading commands (e.g., \texttt{grep}, \texttt{cat}), the raw retrieved context---often thousands of lines---is captured and filtered before reaching the agent. To guide this pruning, the agent generates a \textit{Goal Hint} describing its current information need (e.g., ``Focus on MRO resolution logic''). A lightweight neural skimmer is trained to adaptively selects relevant lines from the raw context, while preserving code structure. %This line-granular approach maintains syntactic validity and semantic coherence while significantly reducing token consumption.
  Only this pruned context (relevant and low-noise) is returned to the agent. %If no Goal Hint is provided, the full raw context is returned, preserving the agent's ability to perform broad exploration when needed.

  The framework comprises three core components as introduced in the following sections, respectively. First, the target agent is instructed to provide a goal hint indicating the current information need (\Cref{subsec:hint}). Next, a lightweight neural skimmer is designed to score context tokens and aggregate relevant lines (\Cref{subsec:model}). %We adopt a compact reranker model architecture and train it with a unified objective combining CRF-based line-level filtering and document-level reranking. The model is deployed with an efficient inference strategy that applies dynamic thresholding. 
  Finally, we integrate the skimmer into real agentic systems with minimal modifications (\Cref{subsec:integration}).% by augmenting tool interfaces with an optional \texttt{context\_focus\_question} parameter, preserving backward compatibility.
  
  \subsection{Goal Hint Generation}
  \label{subsec:hint}
  
  % Central to our framework is the \textit{Goal Hint}---a natural language description of the agent's current information need. \gu{this paragraph does not tell us how the goal hint is generated}\shi{I'll mention the prompt in the appendix}\gu{also need to mention the high-level methodology here} We augment standard file manipulation tools (e.g., \texttt{cat}, \texttt{grep}) with an optional \texttt{context\_focus\_question} parameter\gu{why?}. When provided, this parameter allows agents to express complex reasoning goals \gu{why?} (e.g., \textit{``How is authentication handled?''}) that go beyond simple keyword searches, capturing ``how'' and ``what'' questions about code behavior. When the parameter is omitted, the pruner is bypassed and full outputs are returned, preserving backward compatibility.
  
  Central to our framework is the \textit{Goal Hint}---a natural language description of the agent's current information need. Rather than relying on keyword-based filtering, we instruct the agent to generate goal hints as complete, self-contained questions (e.g., \textit{``How is authentication handled?''}) that capture the semantic intent of its current reasoning step (detailed prompt in~\Cref{app:prompts}). To enable agents to communicate these goal hints to our pruning system, we augment standard file manipulation tools (e.g., \texttt{cat}, \texttt{grep}) with an optional \texttt{context\_focus\_question} parameter. When provided, this parameter passes the goal hint to the skimmer, which then filters large outputs for query-relevant content. When omitted, the skimmer is bypassed and full outputs are returned, preserving backward compatibility. This lightweight wrapper design (illustrated below) requires minimal modifications to existing agent infrastructures, enabling seamless integration without disrupting established workflows.
  
  \begin{lstlisting}[language=Python, style=pythoncode, basicstyle=\small\ttfamily]
  # Original tool (unchanged)
  def grep(file_path, pattern):
      # ... original grep implementation ...
      return matches
  
  # New tool with pruner
  def grep_with_pruner(file_path, pattern, context_focus_question=None):
      # Call original tool
      raw_output = grep(file_path, pattern)  
      if context_focus_question:
          # Prune if hint provided
          return prune(raw_output, context_focus_question)
      return raw_output  # Bypass pruner otherwise
  \end{lstlisting}
  
  \subsection{Lightweight Neural Skimmer}
  \label{subsec:model}
  
  \paragraph{Model Architecture.}
  We formulate context pruning as a reranking problem: Given the agent context $C = \{x_1, x_2, ..., x_n\}$ where each $x_i$ represents a token, and a query $q$ representing the agent's current goal, the model computes a relevance score $s_i$ for each token through a neural scoring function:
  % \wang{A little weird? line is aggregated in inference, but in training stage, we have no only tokens and llm-generated line label will be transformed to token labels, so the emphasize of $l_i$ might be strange and should be replaced with code snippet $c_i$}
  \begin{equation}
      s_i = \mathcal{F}(q, x_i | C; \theta)
  \end{equation}
  where $\mathcal{F}$ represents the scoring function parameterized by $\theta$ that evaluates token $x_i$ in the context of both the query $q$ and the full code context $C$. %This joint encoding ensures that token importance is assessed not in isolation but relative to the specific information need expressed in the query.
  
  To enable line-level pruning decisions, we aggregate token scores to the line level. Let $L = \{l_1, ..., l_m\}$ denote the set of lines obtained by splitting $C$, and let $T_j$ denote the set of tokens in line $l_j$. The line-level relevance score $\bar{s}_j$ is computed as the average of its constituent token scores:
  \begin{equation}
      \bar{s}_j = \frac{1}{|T_j|}\sum_{t \in T_j} s_t
  \end{equation}
  This averaging operation ensures that lines are evaluated based on their overall relevance rather than being dominated by a few high-scoring tokens, maintaining the semantic coherence necessary for code comprehension.
  
  We adopt the Qwen3-Reranker-0.6B~\citep{qwen3embedding} as our backbone due to its efficiency and pre-trained knowledge of code structures. During inference, a line $l_j$ is retained if its aggregated score $\bar{s}_j$ exceeds a predefined threshold $\tau$. The model processes retrieved chunks in parallel to minimize latency. Given its lightweight architecture of 0.6B parameters, pruning overhead is negligible compared to the token savings for downstream agent LLMs. See \Cref{alg:scoring_aggregation} for detailed inference steps.
  
  % \paragraph{Structure-Aware Labeling.}
  % To avoid discarding essential syntactic elements (e.g., closing braces, decorators), we prompt the teacher model to identify both semantically relevant lines and minimal surrounding context needed for syntactic validity. The resulting masks preserve code structure significantly better than token-level compression (\Cref{app:ast_analysis}). We further refine the dataset using an LLM-as-a-Judge filter~\citep{wang2025can,liu2023g,song2024finesure,he2025code,lan2025f2bench,lan2025mcbe} (Qwen3-Next-80B-A3B-Thinking), retaining approximately 1/6 of samples based on reasoning quality.
  
  \paragraph{Training Objective.}
  \label{subsec:training_inference}
  We train the pruning model by minimizing the conditional random field negative log likelihood (CRF-NLL) \citep{zheng2015conditional}. Unlike mere binary cross entropy, CRF explicitly models transition probabilities between retain/prune states, enabling the model to learn line-level retention decisions while capturing sequential dependencies. %This enables the model to learn structured pruning patterns that respect code boundaries. 
  
  Given the context representation $\mathbf{x}=\mathbf{x}_1,\ldots,\mathbf{x}_n$ and silver labels $\mathbf{y}$, the pruning head minimizes the following loss function:
  \begin{equation}
      \mathcal{L}_{\text{compress}} = \frac{1}{B} \sum_{i=1}^{B} \frac{\mathcal{L}_{\text{CRF-NLL}}(\mathbf{x}_i, \mathbf{y}_i)}{L_{i}}
  \end{equation}
  where \(\mathbf{x}_i\) denotes the learned feature representation for each token; $L_i$ is the sequence-length normalization, which prevents bias toward aggressive pruning in long contexts. We keep the original reranking head in Qwen Reranker~\citep{bai2023qwen} to preserve document-level relevance scoring capability. 
  % This dual-head design enables single-pass inference to jointly perform chunk ranking (identifying relevant documents) and intra-chunk pruning (selecting informative lines), which is particularly beneficial for RAG scenarios where both coarse-grained retrieval and fine-grained compression are needed. 
  The reranking head minimizes $\mathcal{L}_{\text{rerank}} = \text{MSE}(s_{\text{pred}}, s_{\text{ref}})$ between predicted and reference document-level relevance scores in $[0,1]$. The final objective combines both heads with a balancing weight $\lambda$:
  \begin{equation}
      \mathcal{L}_{\text{total}} = (1-\lambda) \cdot\mathcal{L}_{\text{compress}} + \lambda \cdot \mathcal{L}_{\text{rerank}}
  \end{equation}
  More details about the architecture, CRF formulations, and training configurations are provided in \Cref{app:architecture}.
  
  \paragraph{Constructing Training Data.}

  Training the neural skimmer requires data with line-level supervision that preserves both relevance and code structure. Since such dataset is unavailable, we construct a polyglot training corpus following a teacher-student paradigm. We sample code snippets from GitHub\footnote{\url{https://huggingface.co/datasets/nick007x/github-code-2025}}, a curated collection of high-quality repositories, and utilize a teacher LLM (Qwen3-Coder-30B-A3B-Instruct~\citep{qwen3technicalreport}) to synthesize task-oriented queries that target specific functional subsets of each snippet. This process produces training quadruplets $(q, C, M, S)$, where $M$ is a binary line-level mask indicating which lines to retain and $S$ is a document-level relevance score. 
  
  To ensure generalization across diverse coding scenarios, we design queries based on a taxonomy of nine distinct agentic tasks, such as \textit{Code Debugging}, \textit{Feature Addition}, and \textit{Code Refactoring}, covering common information needs in real-world agentic workflows. We employ an LLM-as-a-Judge filtering mechanism~\citep{wang2025can, he2025code} with Qwen3-Next-80B-A3B-Thinking~\citep{qwen3technicalreport} to ensure annotation quality. This rigorous filtering yields a final training corpus of 61,184 high-quality samples with verified line-level annotations. Complete details on task taxonomy, data statistics, and generation configurations are provided in \Cref{app:training_data}.

  \subsection{Integration with Agentic Workflows}
  \label{subsec:integration}
  The trained skimmer is deployed in real coding agents to accomplish specific tasks. %Our design follows two key principles: (i) \textit{minimal modification}---the skimmer operates as middleware, requiring minimal changes to existing agent frameworks, and (ii) \textit{task-aware filtering}---agents provide a \textit{Goal Hint} describing their current information need, enabling the skimmer to perform context-aware filtering.
  Our framework flexibly adapts to different task scenarios. For multi-turn agent tasks (e.g., SWE-Bench), agents dynamically generate Goal Hints at each round based on their evolving reasoning trace, enabling them to shift focus from high-level navigation to detailed debugging as needed. For single-turn tasks with inherent queries (e.g., code question answering), the task description serves as the initial Goal Hint, though agents can refine it in subsequent retrieval rounds. This flexibility allows agents to seamlessly transition between broad exploration (no pruning) and focused investigation (with pruning) as their information needs evolve.
  
  \section{Experiments}

  \subsection{Benchmarks and Agents}
  
  We evaluate \approach on four benchmarks spanning single-turn and multi-turn scenarios. For single-turn tasks, we use \textbf{Long Code Completion}~\citep{guo2023longcoder} (500 Python examples with 5K+ token contexts) and \textbf{Long Code QA}~\citep{rando2025longcodebench} (question answering on long code contexts up to 1M tokens), evaluating under 4x and 8x compression constraints. For multi-turn agent tasks, we use \textbf{SWE-Bench Verified}~\citep{jimenez2024swe} (500 real-world GitHub issues requiring patch generation) and \textbf{SWE-QA}~\citep{peng2025swe} (repository-specific question answering across three repositories). Specifically, we integrate \approach into Mini SWE Agent~\citep{yang2024sweagent} for SWE-Bench Verified and OpenHands~\citep{openhands} for SWE-QA, evaluating with Claude Sonnet 4.5 and GLM-4.6 backbone models. Detailed benchmark descriptions, agent configurations, and experimental settings are provided in \Cref{app:experimental_details}.
  
  \subsection{Baselines}
  We compare \approach against representative methods for compressing code context and environment observations. Full Context and No Context establish upper and lower performance bounds. For compression baselines, we evaluate LLMLingua-2~\citep{pan2024llmlingua} and Selective-Context~\citep{li2023compressing} which perform token-level pruning based on perplexity and self-information respectively, RAG which retrieves code chunks via embedding similarity using UniXCoder~\citep{guo2022unixcoder}, and LongCodeZip~\citep{shi2025longcodezip} which leverages program structure for compression. For multi-turn agent tasks, we additionally compare with LLM Summarize that generates abstractive summaries using the backbone model. All baselines are configured to match 4x and 8x compression constraints under identical experimental settings. We did not compare with agent history compression methods~\citep{ye2025agentfold,kang2025acon}. While these methods effectively manage long-horizon agent interactions, they tackle a different problem to ours. They aim at compressing agents' prior interaction trajectories, while our method compresses agents' observations like repository content, making them incomparable to our approach. 
  % Detailed baseline descriptions are provided in \Cref{app:baselines}.
  % \gu{compare with https://arxiv.org/abs/2509.23586?}
  
  \subsection{Metrics}

  We evaluate both task performance and compression efficiency. Task performance is measured through Edit Similarity (ES) and Exact Match (EM) for code completion~\citep{guo2023longcoder}, Accuracy for question answering~\citep{rando2025longcodebench}, Resolve Rate on SWE-Bench Verified~\citep{jimenez2024swe}, and Average LLM-as-a-Judge Score on SWE-QA~\citep{peng2025swe}. Compression efficiency is quantified via Compression Ratio ($1/\tau = |C_{\text{original}}|/|C_{\text{compressed}}|$), absolute Token Consumption, interaction Rounds, and API Cost $(\$)$. 
  % We additionally report \textbf{AST Correctness Rate} to validate structural preservation of compressed code using tree-sitter~\citep{treesitter}. Complete metric definitions and computation details are provided in \Cref{app:metrics}.

  \section{Results}

\subsection{Performance on Multi-Turn Tasks}

% To evaluate \approach in realistic software engineering workflows, we assess its performance when integrated into agent systems. \approach functions as middleware, intercepting file-reading operations and applying task-aware pruning based on dynamically generated focus questions. We conduct experiments on SWE-Bench Verified and SWE-QA, where agents interact with codebases through multiple rounds of exploration and modification.

  \begin{table*}[t]
  \centering
  \label{tab:swebench_agent}
  \resizebox{0.8\textwidth}{!}{%
  \begin{tabular}{lccccc}
  \toprule
  \textbf{Agent} & \textbf{Rounds} & \textbf{Solved} & \textbf{Success (\%)} & \textbf{Tokens (M)} & \textbf{Cost (\$)}\\
  \midrule
Mini SWE Agent (Claude Sonnet 4.5) & 51.0 & 353/500 & 70.6 & 0.911 & 0.504 \\
+ \approach & 41.7 & 360/500 & 72.0\increase{1.4\%} & 0.701\reduction{23.1\%} & 0.369\reduction{26.8\%} \\
\midrule
Mini SWE Agent (GLM 4.6) & 49.3 & 277/500 & 55.4 & 0.791 & 0.055 \\
+ \approach & 36.6 & 283/500 & 56.6\increase{1.2\%} & 0.488\reduction{38.3\%} & 0.035\reduction{36.4\%} \\
  % \midrule
  % Mini SWE Agent(Gemini-3-pro) & 40.3 & 371/500 & 74.2 & 0.677 & 0.487 \\
  % + \approach & 31.7 & 357/500 & 71.4 & 0.655\reduction{3.2\%} & 0.458 \\
  \bottomrule
  \end{tabular}%
  }
  \caption{Results on SWE-Bench Verified. \approach reduces token consumption by 23--38\% while improving success rates by 1.2--1.4 percentage points.}
  \end{table*}

\begin{table}[t]
  \centering
\begin{tabular}{lccc}
\toprule
\textbf{Method} & \textbf{Avg Score} & \textbf{Avg Rounds} & \textbf{Tokens (K)}\\
\midrule
\multicolumn{4}{c}{\textit{Streamlink}}\\
\midrule
Claude Sonnet 4.5 & 8.36 & 23.4 & 611.2 \\
\quad + \approach & \textbf{8.59} \increase{0.23} & 23.9 & 557.1\reduction{8.9\%} \\
\midrule
GLM-4.6 & 8.56 & 18.2 & 318.2\\
\quad + \approach & \textbf{8.56}\increase{0.00} & 25.0 & 145.1\reduction{54.4\%}\\
\midrule
\multicolumn{4}{c}{\textit{Reflex}}\\
\midrule
Claude Sonnet 4.5 & 8.68 & 33.2 & 1081.6 \\
\quad + \approach & \textbf{8.85} \increase{0.17} & 32.4 & 866.8\reduction{19.9\%} \\
\midrule
GLM-4.6 & 8.37 & 26.1 & 142.3\\
\quad + \approach & \textbf{8.23}\decrease{0.14} & 36.7 & 101.2\reduction{28.9\%}\\
\midrule
\multicolumn{4}{c}{\textit{Conan}}\\
\midrule
Claude Sonnet 4.5 & 8.70 & 23.9 & 654.7 \\
\quad + \approach & \textbf{8.84} \increase{0.14} & 23.5 & 520.7\reduction{20.5\%}\\
\midrule
GLM-4.6 & 8.58 & 21.4 & 175.9 \\
\quad + \approach & \textbf{8.45}\decrease{0.13} & 27.7 & 116.6\reduction{33.7\%}\\
\bottomrule
\end{tabular}%

\caption{Results on SWE-QA across different repositories. \approach achieves 29--54\% token reduction on Streamlink, Reflex, and Conan repositories with minimal impact on task performance.}
% \vspace{-0.25cm}
\end{table}

\begin{table}[t]
  \centering
  \setlength{\tabcolsep}{3pt}
\begin{tabular}{lcccc}
\toprule
\textbf{Method} & \textbf{Rounds} & \textbf{Success (\%)} & \textbf{Tokens (M)}\\
\midrule
Mini SWE Agent & 52.3 & \underline{62.0} & 0.972\\
+ LLMLingua2 & 42.1 & 54.0 & 0.856\\
+ RAG & \textbf{40.2} & 50.0 & \underline{0.771}\\
+ LLM Summarize & 41.3 & 56.0 & 0.794\\
+ LongCodeZip & 44.3 & 54.0 & 0.889\\
\rowcolor{blue!10}
+ \textbf{\approach} & \underline{41.1} & \textbf{64.0} & \textbf{0.670}\\
\bottomrule
\end{tabular}%
\caption{Comparison of context compression strategies on SWE-Bench. \approach achieves highest success rate with lowest token usage.}
\label{tab:baseline_comparison}
\end{table}

% We evaluate \approach through cross-model evaluation on coding agent benchmarks (\Cref{tab:swebench_agent,tab:sweqa}) and comparison against alternative context management strategies (\Cref{tab:baseline_comparison}), demonstrating both model-agnostic efficiency and superiority over existing approaches.

\paragraph{Evaluation across Diverse Tasks.}
We integrate \approach into the Mini SWE Agent~\citep{yang2024sweagent} framework for SWE-Bench Verified~\citep{jimenez2024swe} and the OpenHands~\citep{openhands} framework for SWE-QA~\citep{peng2025swe}, evaluating with different backbone models. On SWE-Bench Verified, \approach achieves substantial token reductions of 23--38\% across models while even improving success rates (by 1.2--1.4 percentage points). Notably, interaction rounds decrease by 18--26\%. By filtering redundant information while preserving task-relevant code, \approach enables agents to locate issues more precisely and make more decisive decisions, thereby reducing repeated reads of exploratory files and completing tasks earlier. We conduct in-depth case studies in \Cref{app:case_study} to illustrate these behavioral differences, showing how context pruning transforms both failure-to-success scenarios (with 83.3\% token reduction) and successful trajectories (with 30.2\% reduction in peak prompt length). On SWE-QA, similar efficiency gains emerge with 29--54\% token reduction across repositories. Interestingly, we observe that GLM-4.6 exhibits increased rounds (29--41\% more) while Claude Sonnet 4.5 maintains similar round counts with minor variations (within 3\%). Through trajectory analysis, we find that after pruning, GLM tends to explore more files before formulating answers, suggesting a more conservative reasoning strategy when presented with focused context. Nevertheless, the overall token consumption remains substantially lower, demonstrating that \approach maintains efficiency advantages regardless of the agent's exploration patterns. These model-agnostic efficiency gains directly translate to proportional cost savings. Detailed analysis is provided in \Cref{app:agent_analysis}.

\begin{table*}[t]
  \centering
  % \resizebox{\textwidth}{!}{
  \begin{tabular}{l|rrr|rrr|rr|rr}
  \toprule
  \multirow{3}{*}{Methods}
  & \multicolumn{6}{c|}{Long Code Completion}
  & \multicolumn{4}{c}{Long Code QA} \\
  \cmidrule(lr){2-7}\cmidrule(lr){8-11}
  & \multicolumn{3}{c|}{4x constraint}
  & \multicolumn{3}{c|}{8x constraint}
  & \multicolumn{2}{c|}{4x constraint}
  & \multicolumn{2}{c}{8x constraint} \\
  \cmidrule(lr){2-4}\cmidrule(lr){5-7}\cmidrule(lr){8-9}\cmidrule(lr){10-11}
  & $1/\tau$ & ES & EM
  & $1/\tau$ & ES & EM
  & $1/\tau$ & Acc
  & $1/\tau$ & Acc \\
  \midrule
  
  % \multicolumn{11}{c}{Qwen2.5-Coder-7B-Instruct} \\
  % \midrule
  Full & 1.0 & 64.65 & 40.5 & 1.0 & 64.65 & 40.5 & 1.0 & 54.05 & 1.0 & 54.05 \\
  No Context & $\infty$ & 44.90 & 13.5 & $\infty$ & 44.90 & 13.5 & $\infty$ & 38.39 & $\infty$ & 38.39 \\
  \midrule
  Selective-Context & 3.27 & 52.48 & 22.0 & 7.49 & 48.67 & 17.0 & 3.69 & 55.36 & 7.32 & 51.79 \\
  LLMLingua2 & 3.32 & 49.47 & 15.5 & 7.89 & 44.74 & 13.0 & 3.57 & 55.36 & 7.68 & 51.33 \\
  RAG & 3.29 & 58.97 & 30.5 & 6.60 & 55.82 & 29.0 & 3.06 & 58.04 & 5.87 & 55.86 \\
  LongCodeZip & 2.77 & 57.77 & 28.0 & 7.85 & 56.08 & 27.5 & 3.98 & 52.25 & 7.39 & 54.95 \\
  \rowcolor{blue!10}
  % \textbf{LongCodeZip + \approach} & \textbf{8.26} & \textbf{56.96} & \textbf{29.0} & \textbf{8.26} & \textbf{56.96} & \textbf{29.0} & \textbf{9.00} & \textbf{57.66} & \textbf{30.23} & \textbf{57.27} \\
  \textbf{\approach} & \textbf{5.56} & \textbf{58.63} & \textbf{31.5} & \textbf{10.92} & \textbf{57.58} & \textbf{31.0} & \textbf{13.95} & \textbf{59.46} & \textbf{14.84} & \textbf{58.71} \\
  
  % \midrule
  
  % \multicolumn{11}{c}{Seed-Coder-8B-Instruct} \\
  % \midrule
  % Full & 1.0 & 65.03 & 40.5 & 1.0 & 65.03 & 40.5 & 1.0 & 49.11 & 1.0 & 49.11 \\
  % No Context & $\infty$ & 43.85 & 14.0 & $\infty$ & 43.85 & 14.0 & $\infty$ & 37.50 & $\infty$ & 37.50 \\
  % \midrule
  % Selective-Context & 3.27 & 53.68 & 24.5 & 7.49 & 49.89 & 17.5 & 2.71 & 51.33 & 6.69 & 46.43 \\
  % LLMLingua2 & 3.95 & 48.09 & 15.0 & 7.89 & 45.53 & 13.5 & 3.70 & 43.36 & 5.46 & 39.82  \\
  % RAG & 3.32 & 58.21 & 31.0 & 6.78 & 56.97 & 28.5 & 3.44 & 53.98 & 6.65 & 53.57 \\
  % LongCodeZip & 3.96 & 56.72 & 25.5 & 6.53 & 54.91 & 23.0 & 3.29 & 51.33 & 7.49 & 50.91 \\
  % \rowcolor{blue!10}
  % \textbf{\approach} & \textbf{4.22} & \textbf{57.71} & \textbf{31.0} & \textbf{8.13} & \textbf{56.73} & \textbf{28.5} & \textbf{9.98} & \textbf{56.25} & \textbf{14.68} & \textbf{55.75} \\

  \bottomrule
  \end{tabular}
  % }
  \caption{Main results on Long Code Completion and Long Code QA tasks. The table compares the performance and compression ratio ($1/\tau$) under 4x and 8x constraints. \approach~demonstrates superior effectiveness in maintaining high performance while achieving significant context compression.}
  \label{tab:main-results}
\end{table*}

\paragraph{Comparison with Alternative Context Management Strategies.}
To understand what design choices contribute to \approach's effectiveness, we compare against three categories of baselines: token-level compression (LLMLingua2), coarse-grained retrieval (RAG), and generative summarization (LLM Summarize). Due to computational cost considerations, we conduct this comparison on a random subset with 50 samples of SWE-Bench Verified following~\citet{xia2025live,chen2024coder}. As shown in \Cref{tab:baseline_comparison}, our method achieves the best success rate (64\%), outperforming the vanilla agent baseline (62\%) despite using 31\% fewer tokens. In contrast, LLMLingua2 and RAG degrade performance substantially (to 54\% and 50\%), likely because token-level pruning disrupts code syntax while coarse-grained retrieval misses fine-grained implementation details. LLM Summarize achieves moderate performance (56\%) but incurs additional latency. These results validate our design choice of line-level granularity with task-aware filtering, striking an optimal balance between compression and information retention. The effectiveness of \approach is further validated by consistent improvements on single-turn tasks (\Cref{tab:main-results}), where we observe similar advantages across diverse compression constraints.

\subsection{Performance on Single-Turn Tasks}

While \approach is designed for coding agents, its query-aware, line-level pruning mechanism generalizes to single-turn tasks. We evaluate on Long Code Completion and Long Code QA under 4x and 8x compression constraints with Qwen2.5-Coder-7B-Instruct~\citep{hui2024qwen2}. These benchmarks require direct completion or question answering without iterative exploration. Notably, \approach achieves substantially higher effective compression ratios than baselines configured at identical compression targets. We also validate with Seed-Coder-8B-Instruct~\citep{seed2025seedcoderletcodemodel}, obtaining similar results (see \Cref{app:seedcoder_results}).

On Long Code Completion (\Cref{tab:main-results}), \approach achieves up to 10.92x effective compression under the 8x constraint while maintaining an Edit Similarity (ES) score of 57.58 and an Exact Match (EM) rate of 31.0. As compression constraints tighten, token-level methods (LLMLingua2, Selective-Context) experience rapid performance degradation---Selective-Context drops to 48.67 ES under 8x constraint---whereas \approach maintains stable performance through line-level granularity that preserves syntactic structure. Under the 4x constraint, \approach achieves 5.56x compression with 58.63 ES and 31.5 EM, outperforming all baselines.
On Long Code QA, the advantages become even more pronounced. \approach achieves 14.84x compression under 8x constraint with 58.71\% accuracy, substantially exceeding other baselines. Under the 4x constraint, our method achieves 13.95x compression while maintaining 59.46\% accuracy, demonstrating that task-aware line-level pruning excels at identifying relevant code segments for question answering. This highlights the synergy between query-aware filtering and line-level granularity, validating that our approach generalizes effectively to single-turn code understanding tasks.

% \wang{TODO:For detail 5-dim score, we might need a Radar chart?} \shi{We can omit this since it's just sub-metrics for a sub-task we are testing on}

\subsection{Efficiency Impact}
\label{subsec:efficiency_impact}

\begin{figure}[t]
    \centering
    \includegraphics[width=0.4\columnwidth]{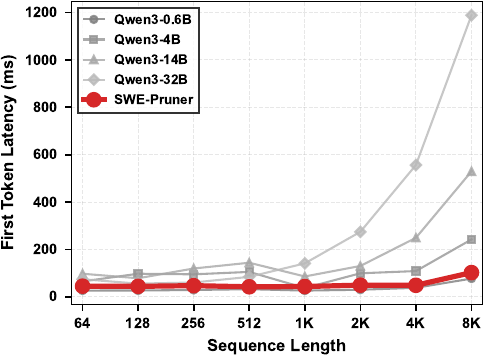}
    \caption{First token latency comparison across different sequence lengths. \approach maintains consistently low latency below 100 ms.}
    % \vspace{-0.25cm}
    \label{fig:ttft_comparison}
\end{figure}

Beyond token reduction, a critical consideration for practical deployment is the computational overhead introduced by the skimmer itself. As shown in \Cref{fig:ttft_comparison}, our approach maintains remarkably low and stable first token latency across all sequence lengths, remaining below 100ms even at 8K tokens. In contrast, larger generative models exhibit exponential latency growth, with Qwen3-32B exceeding 1200ms and closed-source models incurring even more overhead. Since our skimmer employs only a 0.6B encoder, its computational cost is negligible and readily amortized by the reduced decoding cost from context compression. 
% Furthermore, the compact model size of 0.6B requires minimal GPU memory footprint, making deployment highly accessible across diverse hardware configurations. 
This lightweight characteristic validates the practical viability of our approach for real-world applications. Detailed latency measurements are provided in \Cref{tab:ttft_results}.

% We utilize fvcore \wang{add citation, https://github.com/facebookresearch/fvcore} to calculate the FLOPs of model inference. For a 1024-token input, our method requires 0.5 TFLOPs, equivalent to a forward pass of the Qwen3/Reranker-0.6B backbone model. This computational cost is substantially lower than that of current mainstream coding agent models, rendering our approach nearly cost-free compared to expensive generative compression methods.
% $$C_{prefill} \approx \underbrace{2 \cdot N \cdot s}_{\text{(Linear Layers)}} + \underbrace{2 \cdot L \cdot s^2 \cdot d_{model}}_{\text{(Attention)}}$$
% 37B -> 37T（use fp16）
% \subsection{Multi-lingual Generalization}
% Single-language and joint models are evaluated on Python, Java, Go, C/C++, JavaScript, and Markdown. Or using SWE-Bench Multilingual. \shi{TBD.}
% \wang{May be have not enough time}

\section{Related Works}
% \shi{TODO: discuss about works on agentic context folding, even in another discussion section}
\paragraph{Prompt Compression}
Prompt compression has been extensively studied for both natural language and code. Token-level pruning methods such as LLMLingua~\citep{jiang2023llmlingua,jiang2024longllmlingua}, Selective Context~\citep{li2023compressing}, and AttentionRAG~\citep{fang2025attentionrag}, along with embedding/retrieval-based approaches like RAG~\citep{lewis2020retrieval,cheng2026resolvingrobustnessprecisiontradeofffinancial,shi2026reasoning}, XRAG~\citep{cheng2024xrag}, and repo-level retrieval for code completion~\citep{zhang2023repocoder,lai2026transformers}, can reduce prompt length on single-round tasks but often fail to preserve syntactic structures essential for code correctness~\citep{shi2025longcodezip}. Code-specific methods including DietCode~\citep{zhang2022diet}, SlimCode~\citep{wang2024natural}, LongCodeZip~\citep{shi2025longcodezip}, CodeOCR~\citep{shi2026codeocr}, and context inlining~\citep{hu2026line,zeng2025pruning} address structural concerns but are primarily evaluated on single-round proxy tasks (e.g., code search, completion) rather than multi-round agentic workflows. Moreover, these approaches typically apply static, content-only compression that lacks contextual awareness, leading to fixed behaviors such as indiscriminately removing comments~\citep{yang2024less} regardless of task requirements. In contrast, \approach~performs line-level pruning with dynamic, query-conditioned thresholding that adapts to the agent's current task stage, enabling context-aware compression validated on end-to-end benchmarks like SWE-Bench~\citep{jimenez2024swe} without repository-specific retraining.

\paragraph{Agent Context Management}
Modern coding LLMs support context windows of 128k tokens or more~\citep{achiam2023gpt,team2024gemini}, yet they remain insufficient for large-scale codebases and suffer from documented performance degradation on long contexts~\citep{liu2023lost,laban2025llms,li2023loogle}. Agentic systems~\citep{yao2023react,wei2022chain} for software engineering~\citep{yang2024sweagent,wang2024openhands,xia2024agentless,bouzenia2024repairagent,qin2024agentfl,liu2024large,fang2025lastingbench,zhao2026dllm} face critical challenges in managing multi-turn interaction contexts. In NLP domains, trajectory management approaches employ LLM-based summarization when contexts overflow~\citep{cursor,claude_code}, fixed-size truncation~\citep{gao2025trae}, or simple observation masking~\citep{lindenbauer2025complexity}. For long-horizon scenarios, recent frameworks such as SUPO~\citep{lu2025supo} and FoldGRPO~\citep{sun2025contextfolding} learn to manage history through reinforcement learning, while others like COMPASS~\citep{wan2025compass}, ACON~\citep{kang2025acon}, AgentDiet~\citep{xiao2025improving}, and AgentFold~\citep{ye2025agentfold} introduce hierarchical oversight or proactive folding policies. These methods primarily optimize prior interaction histories, whereas \approach~serves as lightweight middleware at the agent--environment boundary to prune the environment's observation (i.e., file content); it is thus orthogonal to and can be seamlessly combined with such learned history managers.

\section{Conclusion}

Pruning long code context for agents requires both fine-grained, structure-preserving compression and dynamic, task-aware filtering. Our approach, \approach, addresses these challenges through lightweight binary classifiers trained on synthetically diversified code-question pairs and an adaptive, query-conditioned thresholding mechanism. By integrating seamlessly as middleware within agentic workflows, \approach reduces token usage by 23--38\% on SWE-Bench while improving success rates by 1.2--1.4 percentage points, achieves 29--54\% reduction on SWE-QA, and up to 14.84$\times$ compression on single-turn benchmarks. These results demonstrate that line-level, context-aware pruning effectively addresses context-window constraints across both agentic workflows and general code understanding tasks while reducing costs and improving efficiency.

\section*{Limitations}

We acknowledge several limitations. First, our implementation focuses on Python repositories, though our approach does not rely on Python-specific features and demonstrates effective generalization across different codebases. Comprehensive multilingual support remains future work. Second, we mitigate data leakage by selecting recently collected repositories from SWE-QA that postdate our training data, though continuous evaluation on newly released repositories remains important. Finally, while our lightweight neural skimmer significantly reduces token consumption, it introduces marginal latency overhead that could be further optimized through distillation or early-exit mechanisms. 
% Finally, our approach currently focuses on code compression without integrating multi-modal signals such as documentation or repository structure.

\section*{Acknowledgments}
This paper is supported by the National Key Research and Development Program of China (Grant No. 2023YFB4503802) and the Natural Science Foundation of Shanghai (Grant No. 25ZR1401175).

% \clearpage
\bibliographystyle{plainnat}
\bibliography{reference}

% % \clearpage
% \input{text/appendix}

% \clearpage
\appendix

\section{Empirical Results on GLM Model}
\label{app:glm_analysis}

In our main experiments, we demonstrated that Claude Sonnet 4.5 dedicates 76.1\% of its token budget to read operations when solving software engineering tasks on SWE-Bench Verified. To verify whether this token consumption pattern generalizes across different model architectures, we extend our analysis to GLM-4.6, an open-source large language model with fundamentally different training methodology and architecture. We conduct parallel analysis on GLM-4.6 agent trajectories using the same experimental setup as in \Cref{sec:approach}, categorizing token usage into read, edit, and execute operations. The results reveal consistent patterns that validate the model-agnostic nature of context pruning needs in coding agents.

\begin{figure}[htbp]
  \centering
  \includegraphics[width=0.4\linewidth]{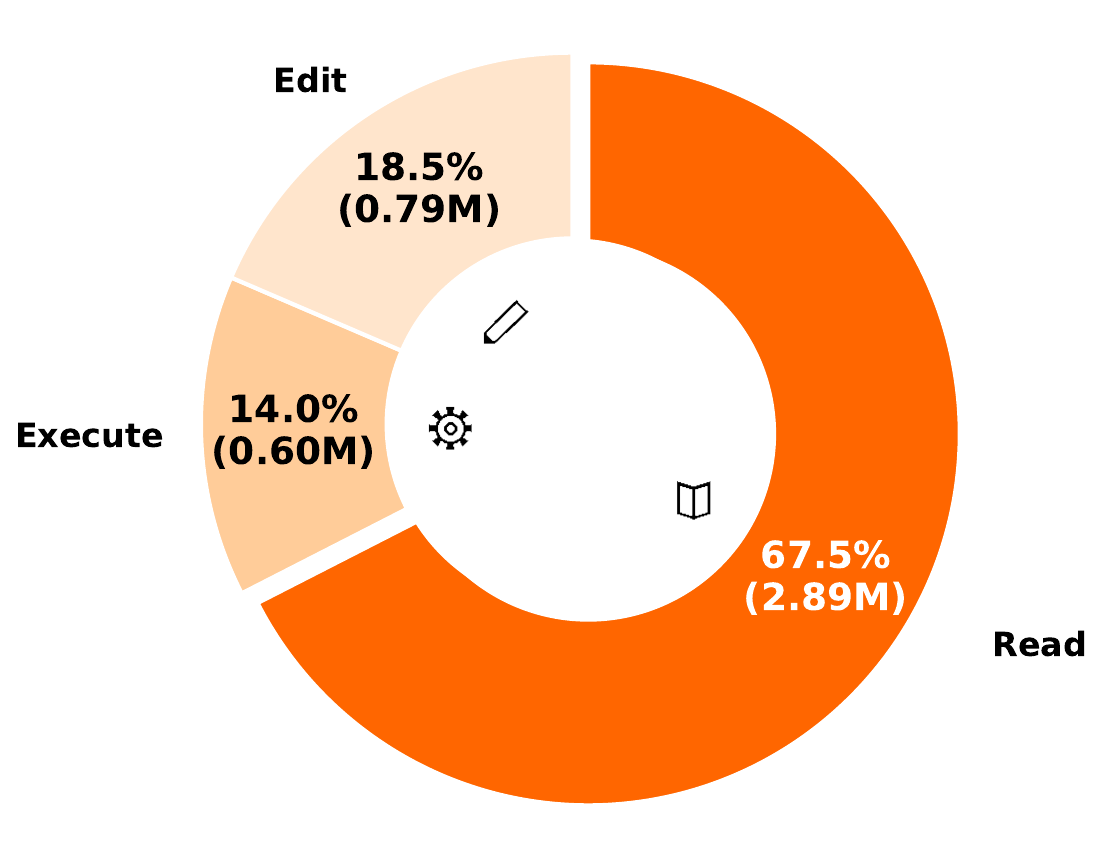}
  \caption{\textbf{Token cost distribution} over different tool calls for Mini-SWE-Agent with GLM-4.6 on SWE-Bench Verified. Read operations dominate token consumption at 67.5\%, further validating the necessity of context pruning mechanisms across different backbone models.}
  \label{fig:glm_token_count}
\end{figure}

As illustrated in \Cref{fig:glm_token_count}, GLM-4.6 exhibits remarkably similar token consumption patterns to Claude Sonnet 4.5. Read-type operations account for 67.5\% of total tokens (2.89M), demonstrating that the dominance of codebase exploration remains consistent across different model architectures. Edit and execute operations consume 18.5\% (0.79M) and 14.0\% (0.60M) respectively, with their relative proportions slightly different from Claude Sonnet 4.5 but still maintaining read operations as the overwhelming majority. This cross-model consistency strongly reinforces our motivation for context pruning: regardless of the underlying model architecture, training methodology, or parameter scale, coding agents universally spend the majority of their token budget on codebase exploration rather than reasoning or editing. These findings establish that \approach addresses a fundamental inefficiency inherent to the agentic workflow itself, rather than model-specific behavior.

\section{Model Architecture and Inference Details}
\label{app:architecture}

\begin{figure*}[t]
  \centering
  \includegraphics[width=0.9\linewidth]{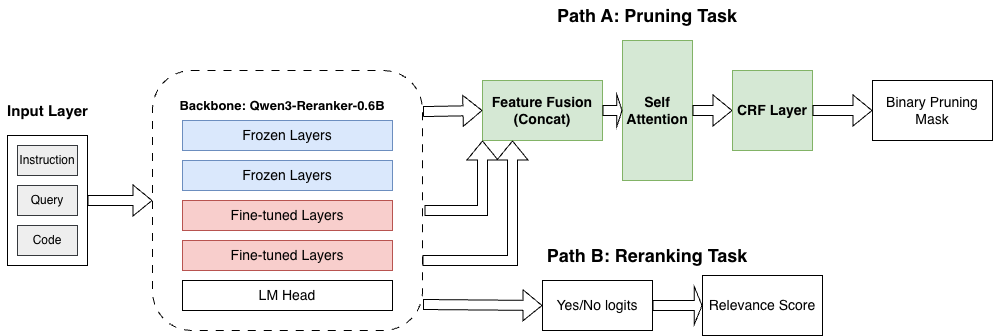}
  \caption{\textbf{\approach Model Architecture.} The model consists of a lightweight reranker backbone with multi-layer feature fusion, followed by dual heads for pruning and reranking. The pruning head employs a CRF layer to model structured retention decisions, while the reranking head produces document-level relevance scores.}
  \label{fig:model-arch}
\end{figure*}

\subsection{Model Architecture}

The neural skimmer extends the Qwen3-Reranker-0.6B backbone with two specialized heads: a CRF-based pruning head for line-level filtering and a reranking head for document-level scoring. An overview of the architecture is shown in \Cref{fig:model-arch}.

\paragraph{Multi-Layer Feature Fusion.} To capture representations at different semantic levels, we extract and concatenate hidden states from three intermediate layers (layers 7, 14, and 28) of the backbone. These fused features are processed through a self-attention block followed by a multi-head attention layer (8 heads, hidden size 256) to refine contextual representations.

\paragraph{CRF-Based Pruning Head.} The pruning head formulates line-level retention as a structured sequence labeling problem using a Conditional Random Field. Let $\mathbf{x}$ denote the input token sequence and $\mathbf{y}$ the corresponding label sequence in space $\mathcal{Y} = \{\text{retain}, \text{prune}\}$. The CRF negative log-likelihood is:
\begin{equation}
    \mathcal{L}_{\text{CRF-NLL}}(\mathbf{x}, \mathbf{y}) = \log Z(\mathbf{x}) - \text{score}(\mathbf{x}, \mathbf{y})
\end{equation}
where the score function combines emission and transition potentials:
\begin{dmath}
    \text{score}(\mathbf{x}, \mathbf{y}) = \text{start}_{y_1} + \sum_{t=1}^{T} \text{emissions}_{t, y_t} + \sum_{t=2}^{T} \text{transitions}_{y_t, y_{t-1}} + \text{end}_{y_T}
\end{dmath}
and the partition function normalizes over all possible label sequences:
\begin{equation}
    \log Z(\mathbf{x}) = \log \sum_{\mathbf{y}' \in \mathcal{Y}} \exp(\text{score}(\mathbf{x}, \mathbf{y}'))
\end{equation}
Emissions $\mathbf{E}_t = \text{MLP}(\mathbf{h}_t) \in \mathbb{R}^2$ represent local confidence for each token, while transitions $\mathbf{T} \in \mathbb{R}^{2 \times 2}$ capture dependencies between adjacent decisions. This structured formulation encourages coherent pruning patterns that respect syntactic boundaries.

\paragraph{Reranking Head.} The reranking head reuses the original language modeling head from Qwen3-Reranker, producing a scalar relevance score for the entire document via mean squared error against teacher-provided scores.

\subsection{Inference Strategy}

During inference, the pruning head computes token-level scores through forward propagation, which are then aggregated to line-level scores via averaging (as detailed in \Cref{alg:scoring_aggregation}). The CRF layer applies Viterbi decoding to find the optimal label sequence, ensuring structurally coherent pruning decisions. Lines are retained if their average token score exceeds the threshold $\tau=0.5$. The reranking head simultaneously produces document-level scores, enabling the system to perform both granular pruning and coarse-grained relevance assessment in a single forward pass.

\begin{algorithm}[t]
\caption{Token Scoring and Line-level Aggregation}
\label{alg:scoring_aggregation}
\begin{algorithmic}[1]
\REQUIRE Input text $X = \{x_1, x_2, ..., x_n\}$, token-level scoring model $f(\cdot)$, threshold $\tau$
\ENSURE Set of kept lines $L_{kept}$
\STATE $L_{kept}=\emptyset$
\STATE Compute token scores $S = \{s_1, s_2, ..., s_n\}$ where $s_i = f(x_i)$
\FOR{each line $l_i \in X$}
    \STATE Let $T_i$ be the set of tokens in line $l_i$
    \STATE Compute line score $\bar{s}_i = \frac{1}{|T_i|}\sum_{t \in T_i} s_t$
    \IF{$\bar{s}_i > \tau$}
        \STATE $L_{kept} \leftarrow L_{kept} \cup \{l_i\}$
    \ENDIF
\ENDFOR
\RETURN $L_{kept}$
\end{algorithmic}
\end{algorithm}

\Cref{alg:scoring_aggregation} summarizes the complete scoring and aggregation pipeline. The algorithm first computes token-level relevance scores for all input tokens (Step 2), then iterates through each line to aggregate token scores via averaging (Steps 4-7), and finally applies the threshold-based retention criterion (Step 8) to produce the final set of kept lines.

% ===============================================================
% Part II: Experimental Details
% ===============================================================

\section{Training Dataset for the Neural Skimmer}
\label{app:training_data}

% \wang{before filtering: 200000 samples from 195370 files from 5945 repos}
% \wang{query generation: temperature 0.7, top p 0.9}
% after filtering and enhancement:  61184
% query word cnt: avg 39.98, medium 24.00
% query char cnt: avg 291.69, medium 169.00
% score 分布暂时不放了？基本都是0 1

\paragraph{Code Source and Preprocessing.}
We construct our training dataset from code snippets sampled from the GitHub Code 2025 dataset\footnote{\url{https://huggingface.co/datasets/nick007x/github-code-2025}}, a meticulously curated collection comprising over 1.5 million repositories. This dataset employs a dual-perspective design that balances proven quality with emerging innovation: it includes both high-quality repositories (above 2 stars) representing established patterns and practices, as well as newly-created 2025 repositories capturing contemporary development trends. The dataset undergoes extensive preprocessing to remove binary files, build artifacts, configuration noise, and minified code, ensuring that only clean, meaningful source code is retained. This rigorous curation provides us with a diverse, polyglot code corpus spanning multiple programming languages and coding paradigms, which is essential for training a pruning model that generalizes across varied agentic scenarios.

\paragraph{Agentic Task Taxonomy.}
To ensure the skimmer generalizes across diverse real-world coding scenarios, we design a comprehensive taxonomy of nine distinct agentic task types that reflect common information needs in software development workflows. These tasks encompass both exploratory activities and focused interventions. \textit{Code Summarization} (\texttt{code-summarize}) targets high-level understanding by requesting concise summaries of code functionality for integration or review purposes. \textit{Code Refactoring} (\texttt{code-refactor}) focuses on improving code quality through readability, modularity, or structural enhancements. \textit{Relevant Part Identification} (\texttt{find-relevant-part}) and \textit{Code Location} (\texttt{code-locate}) address navigation needs by asking to identify where specific features, bugs, or logic are implemented. \textit{Code Optimization} (\texttt{code-optimize}) requests efficiency improvements in terms of performance, resource usage, or scalability. \textit{Code Explanation} (\texttt{code-explain}) seeks understanding of particular algorithms or design choices without requiring full code walkthroughs. \textit{Code Debugging} (\texttt{code-debug}) targets actionable assistance for resolving specific issues, exceptions, or edge cases. \textit{Feature Addition} (\texttt{feature-addition}) involves extending existing code with new capabilities while maintaining integration with current logic. Finally, \textit{Code Completion} (\texttt{code-completion}) represents a unique scenario where the query itself is a code snippet requiring contextual completion, simulating realistic code intelligence workflows. This taxonomy ensures broad coverage of task diversity while maintaining clear distinctions between query intents. Detailed generation instructions for each task type are provided in \Cref{tab:query_tasks}.

\paragraph{Data Generation Pipeline.}
We synthesize queries and line-level labels for 200,000 sampled code snippets from 195,370 files across 5,945 repositories. For each code snippet $C$, we employ Qwen3-Coder-30B-A3B-Instruct~\citep{bai2023qwen} as the teacher LLM to generate task-oriented queries and corresponding line-level retention masks. Query generation employs a temperature of 0.7 and top-p sampling with $p=0.9$ to balance diversity and coherence. To ensure representativeness, we randomly sample across all nine task types, three snippet length levels (short, medium, long), and three relevance levels (low, medium, high). Following initial generation, we apply a rigorous quality control mechanism using Qwen3-Next-80B-A3B-Thinking as an LLM-as-a-Judge filter~\citep{wang2025can,liu2023g,song2024finesure,he2025code,lan2025f2bench,lan2025mcbe}. This judge model evaluates the reasoning quality, annotation consistency, and task alignment of each sample, retaining approximately 1/6 of candidates that meet high-quality standards. The complete generation prompts and judge criteria are detailed in \Cref{app:prompts}.

\paragraph{Dataset Statistics.}
Following filtering and quality enhancement procedures, we obtain 61,184 training samples with verified line-level annotations. The resulting dataset exhibits natural query length distributions with an average of 39.98 words and a median of 24.00 words per query, reflecting realistic information needs in agentic workflows. The average character count per query is 291.69, with a median of 169.00, indicating a balanced mix of concise and detailed information requests. This final corpus provides diverse, high-quality supervision for training the neural skimmer to handle varied agentic coding scenarios.

\begin{table*}[t]
\centering
\caption{Agentic Tasks Taxonomy used for Query Synthesis}
\label{tab:query_tasks}
\begin{tabularx}{\textwidth}{lX}
\toprule
\textbf{Task Type} & \textbf{Instruction for Query Generation} \\
\midrule
\texttt{code-summarize} & Summarize the main purpose or functionality of the code, but do not explain every line. Frame your query as a developer seeking a summary for integration or review. \\
\midrule
\texttt{code-refactor} & Suggest a refactoring or improvement for the code. Your query should be practical, such as asking to improve readability, modularity, or performance. \\
\midrule
\texttt{find-relevant-part} & Ask to locate or identify the part of the code that implements a specific feature or logic. Your query should be about finding where something is handled in the code. \\
\midrule
\texttt{code-optimize} & Request an optimization for the (core logic maybe) code, such as improving efficiency, reducing resource usage, or enhancing scalability. \\
\midrule
\texttt{code-locate} & Ask to pinpoint the location of a bug, feature, or important logic within the code. \\
\midrule
\texttt{code-explain} & Request an explanation for a particular logic, algorithm, or design choice in the code, but do not ask for a full code walkthrough. \\
\midrule
\texttt{code-debug} & Ask for help debugging a specific issue, exception, or edge case in the code. Your query should be actionable and focused. \\
\midrule
\texttt{feature-addition} & Request to add a new feature or capability to the code, specifying what should be added and how it should interact with existing logic. \\
\midrule
\texttt{code-completion} & This is a special query format. In code completion, the query should be CODE instead of text, which means you should image yourself as a developer write other code snippet(query) that can used the code given for completion. The completion will be the next line for query, but you should keep it in your mind and never write the completion in query. QUERY like a PUZZLE. \\
\bottomrule
\end{tabularx}
\end{table*}

\section{Experimental Details}
\label{app:experimental_details}

This section provides comprehensive details about our experimental setup, including training hyperparameters, benchmark specifications, agent frameworks, and baseline configurations.
% This section provides comprehensive details about our experimental setup, including hardware configuration, training hyperparameters, benchmark specifications, agent frameworks, and baseline configurations.

\subsection{Training Configuration}
\label{app:hardware_training}

% \paragraph{Hardware.} All model training is conducted on a server equipped with 8 NVIDIA H100-80G GPUs. The lightweight neural skimmer (0.6B parameters) is deployed on a single GPU for inference, introducing negligible latency overhead compared to the main agent LLM.

\paragraph{Training Hyperparameters.} We fine-tune the Qwen3-Reranker-0.6B backbone with a global batch size of 128, obtained from a per-device batch size of 16 on 8 GPUs with tensor parallelism. We use the AdamW optimizer with a learning rate of $3\times10^{-5}$ and weight decay of 0.01, training for 3 epochs with a dropout rate of 0.4. The CRF-based pruning head fuses hidden states from three intermediate transformer layers (layers 7, 14, and 28). Only the last two transformer layers of the backbone are fine-tuned, along with an additional feature fusion module consisting of a self-attention block followed by multi-head attention (8 heads, hidden size 256) and a CRF layer. The weight balancing parameter is set to $\lambda=0.05$. Detailed architecture specifications are provided in \Cref{app:architecture}.

\paragraph{Inference Configuration.} The pruning threshold is set to $\tau=0.5$, tuned on a held-out validation set. Agent tasks use a maximum of 250 interaction rounds. All experiments employ Claude Sonnet 4.5 and GLM-4.6 APIs with temperature 0 for deterministic generation, averaged over three random seeds where applicable.

\subsection{Benchmark and Agent Descriptions}
\label{app:benchmarks}

We evaluate \approach on both single-turn and multi-turn benchmarks spanning diverse code intelligence tasks. For single-turn evaluation, we use Long Code Completion~\citep{guo2023longcoder}, which evaluates code completion under long contexts, and Long Code QA~\citep{rando2025longcodebench}, which tests code comprehension through question answering on contexts up to 1 million tokens drawn from real-world GitHub issues and documentation across multiple languages and project domains. For multi-turn agent benchmarks, we use SWE-Bench Verified~\citep{jimenez2024swe}, which contains 500 GitHub issues from 12 Python repositories where success is measured by automated test execution in Docker containers and patches must pass all tests without regressions, and SWE-QA~\citep{peng2025swe}, which evaluates repository-specific question answering across three repositories (streamlink, reflex, conan) with answers scored via LLM-as-a-judge across five dimensions: correctness, completeness, relevance, clarity, and reasoning.

For agent frameworks, we integrate \approach with two representative systems. Mini SWE Agent~\citep{yang2024sweagent} is a typical agent framework designed for SWE-Bench that operates with three tool categories: file reading (\texttt{cat}, \texttt{grep}), code editing (\texttt{sed}), and command execution. We integrate \approach as middleware intercepting file reading operations to apply task-aware pruning. OpenHands~\citep{openhands}, which we use following the SWE-QA original paper, provides comprehensive tools including repository exploration, version control, and debugging. Our integration with both frameworks demonstrates that \approach generalizes across agent architectures as a modular component.

\subsection{Baseline Configurations}
\label{app:baseline_configs}

\begin{figure*}[t]
  \centering
  \subfloat[Claude Sonnet 4.5]{\includegraphics[width=0.45\textwidth]{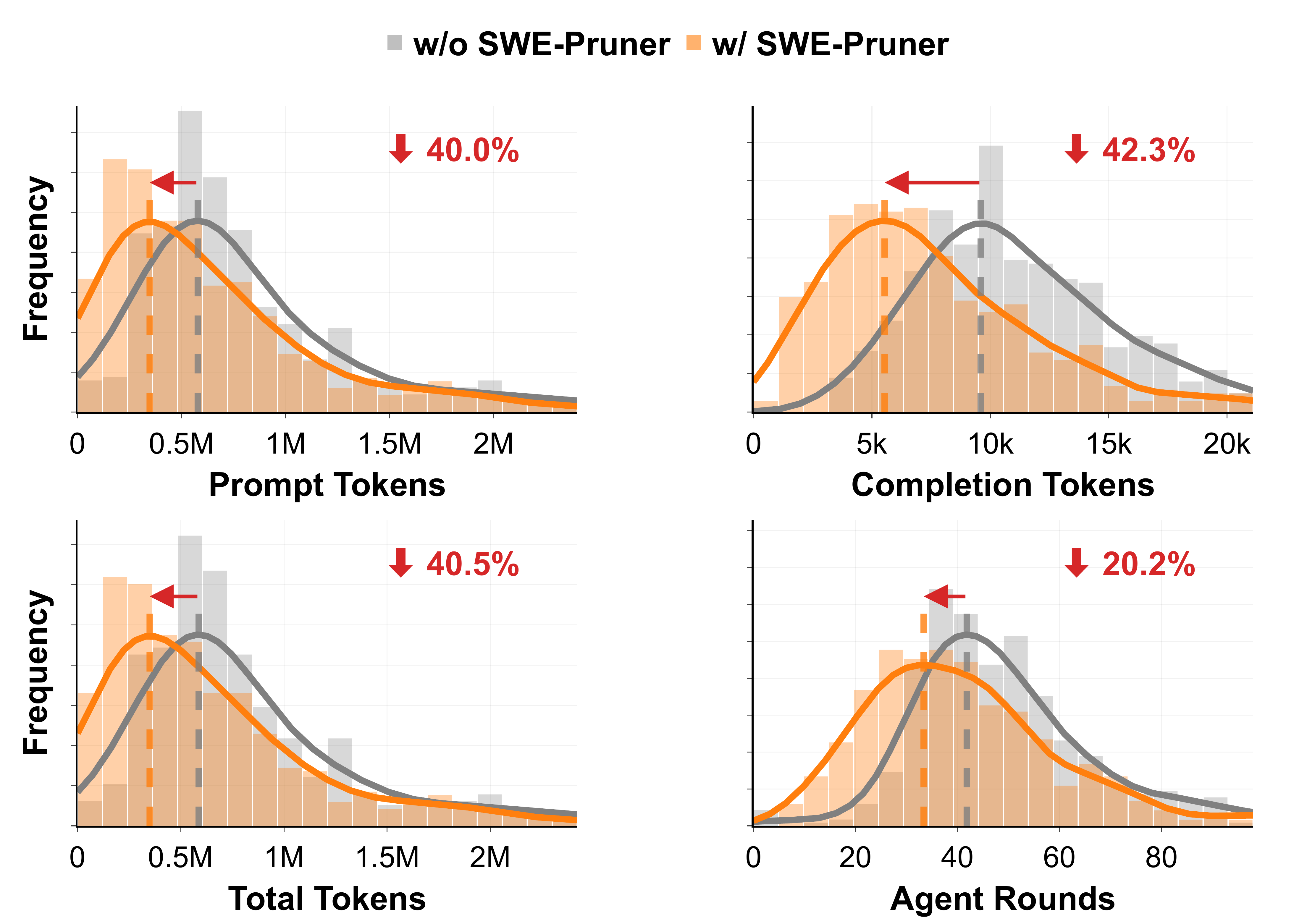}\label{fig:claude_compare_full}}
  \subfloat[GLM 4.6]{\includegraphics[width=0.45\textwidth]{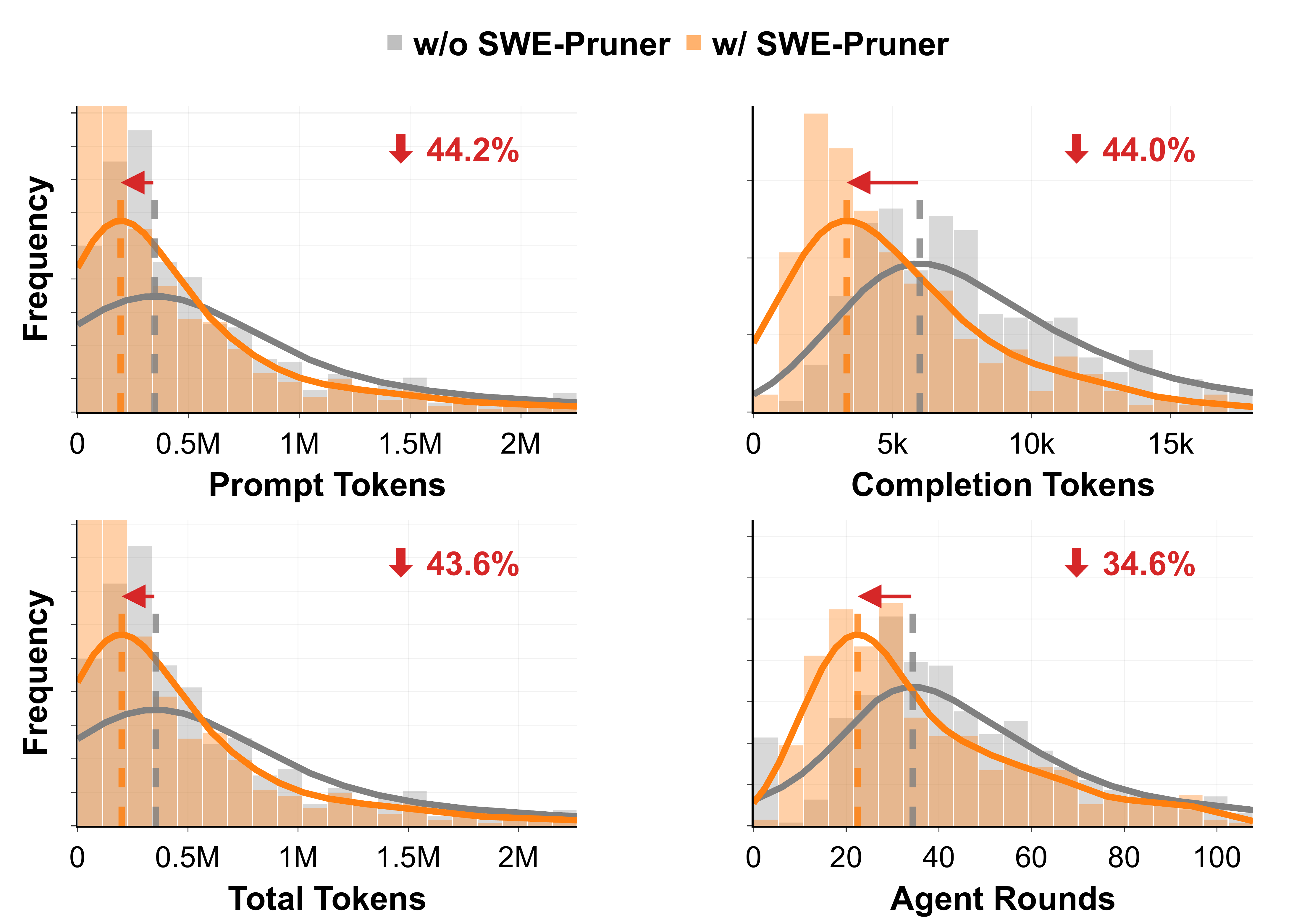}\label{fig:glm_compare_full}}
  \hfill
  \caption{\textbf{Comprehensive efficiency analysis on SWE-Bench Verified.} \approach (\textcolor{orange}{orange}) achieves substantial reductions compared to baseline (\textcolor{gray}{gray}). (a) With Claude Sonnet 4.5: 38.7\% in prompt tokens, 40.8\% in completion tokens, 39.2\% in total tokens, and 18.3\% in agent rounds. (b) With GLM 4.6: 44.2\% in prompt tokens, 44.0\% in completion tokens, 43.6\% in total tokens, and 34.6\% in agent rounds. }
  \label{fig:claude-glm-compare-full}
\end{figure*}

All baseline methods are configured to match identical compression constraints (4x and 8x) under the same experimental conditions. For performance bounds, \textbf{Full Context} provides complete code context as an upper bound, while \textbf{No Context} provides only task instructions as a lower bound. For token-level compression baselines, \textbf{Selective-Context}~\citep{li2023compressing} computes self-information $-\log P(x_i | \text{instruction})$ and removes low-information tokens, and \textbf{LLMLingua-2}~\citep{pan2024llmlingua} uses a trained token classifier (XLM-RoBERTa) to predict binary retention decisions. Both operate at sub-word granularity without code-specific syntactic constraints. For retrieval-based methods, \textbf{RAG} employs function-level chunking with UniXCoder~\citep{guo2022unixcoder,zhang2023repocoder} embeddings, retrieving top-k chunks via cosine similarity. \textbf{LongCodeZip}~\citep{shi2025longcodezip} represents code-specific compression that combines AST-based chunking with entropy-guided compression, retaining high-entropy regions. Finally, \textbf{LLM Summarize} (for multi-turn agent tasks only) generates abstractive summaries of code files using the backbone model, trading summarization cost for reduced downstream tokens.

\section{Agent Rounds and Token Consumption Analysis}
\label{app:agent_analysis}

\Cref{tab:swebench_agent} shows the token consumption and interaction rounds for two different backbone models: Claude Sonnet 4.5 and GLM 4.6. For Claude Sonnet 4.5, \approach reduces average tokens per instance from 0.911M to 0.701M (23.1\% reduction), while reducing average rounds from 51.0 to 41.7 (18.2\% reduction). The GLM 4.6 model exhibits even more substantial improvements, with token reduction from 0.791M to 0.488M (38.3\% reduction) and rounds reduction from 49.3 to 36.6 (25.7\% reduction).

\Cref{fig:claude-glm-compare-full} present the complete distribution shifts for both models, revealing consistent patterns across prompt and completion tokens. GLM 4.6 exhibits 44.2\% and 44.0\% reductions respectively, while Claude Sonnet 4.5 demonstrates 38.7\% and 40.8\% reductions. This symmetry indicates that \approach not only reduces input context but also enables more focused agent responses, distinguishing it from naive retrieval filtering which would primarily affect prompt tokens. The variation in pruning effectiveness between models provides insights into architectural differences: GLM 4.6's more pronounced gains (34.6\% round reduction vs. Claude's 18.3\%) suggest greater susceptibility to context noise, while Claude's extensive context capabilities maintain reasonable focus even unpruned. The reduction in agent rounds represents a qualitative improvement—for GLM 4.6, the shift from 49.3 to 36.6 rounds translates to faster task completion and reduced cumulative latency in production environments where each round incurs API overhead and rate-limiting delays.

\section{Detailed Efficiency Analysis}
\label{app:efficiency}

\begin{table}[htbp]
  \centering
  % \resizebox{\columnwidth}{!}{
  \begin{tabular}{lccccc}
  \toprule
  \textbf{Model} & \multicolumn{5}{c}{\textbf{Input Length}} \\
  \cmidrule(lr){2-6}
               & \textbf{64} & \textbf{128} & \textbf{512} & \textbf{2048} & \textbf{8192} \\
  \midrule
  Qwen3-0.6B     & 26.18 & 26.31 & 32.00 & 29.64  & 76.73  \\
  Qwen3-4B       & 64.75 & 97.24 & 104.93& 99.10  & 241.97 \\
  Qwen3-14B      & 97.17 & 78.11 & 143.65& 129.97 & 529.45 \\
  Qwen3-32B      & 73.99 & 55.46 & 84.01 & 274.22 & 1188.67\\
  \rowcolor{blue!10}
  SWE-Pruner     & 44.70 & 43.91 & 42.05 & 49.05  & 102.00 \\
  \bottomrule
  \end{tabular}
  % }
  \caption{Average TTFT (ms) for different models and input lengths. \approach maintains consistently low latency across all sequence lengths.}
  \label{tab:ttft_results}
\end{table}

To complement the efficiency analysis in Section~\ref{subsec:efficiency_impact}, we provide detailed first token latency measurements across different models and input lengths. \Cref{tab:ttft_results} presents a comprehensive comparison demonstrating how \approach maintains stable and low latency across various sequence lengths compared to larger generative models.

The latency analysis reveals critical insights for practical agent deployments. At 8192 tokens, \approach achieves 102.00ms TTFT—a 7.5$\times$ speedup compared to Qwen3-32B (1188.67ms) and 5.2$\times$ compared to Qwen3-14B (529.45ms)—while exhibiting sublinear scaling (2.1$\times$ increase from 2048 to 8192 tokens vs. Qwen3-32B's 14.1$\times$). Crucially, in real-world deployments with closed-source models like Claude Sonnet 4.5, API latency typically ranges from 500ms to several seconds per request, making our pruning overhead (40--50ms) less than 10\% of typical roundtrip times. This marginal cost is amortized many times over through 23--54\% token reductions (\Cref{tab:swebench_agent}), yielding proportional savings in both inference time and API costs. Combined with 18.3--25.7\% reductions in interaction rounds across models, the efficiency gains compound throughout multi-turn agent trajectories, delivering substantial end-to-end improvements despite the upfront pruning cost.

\section{Single-Turn Tasks with SeedCoder}
\label{app:seedcoder_results}

To validate the generalizability of our approach across different model architectures, we conduct additional evaluations using Seed-Coder-8B-Instruct on the Long Code Completion and Long Code QA benchmarks. Results are presented in \Cref{tab:seedcoder-results}.

\begin{table*}[t]
  \centering
  \begin{tabular}{l|rrr|rrr|rr|rr}
  \toprule
  \multirow{3}{*}{Methods}
  & \multicolumn{6}{c|}{Long Code Completion}
  & \multicolumn{4}{c}{Long Code QA} \\
  \cmidrule(lr){2-7}\cmidrule(lr){8-11}
  & \multicolumn{3}{c|}{4x constraint}
  & \multicolumn{3}{c|}{8x constraint}
  & \multicolumn{2}{c|}{4x constraint}
  & \multicolumn{2}{c}{8x constraint} \\
  \cmidrule(lr){2-4}\cmidrule(lr){5-7}\cmidrule(lr){8-9}\cmidrule(lr){10-11}
  & $1/\tau$ & ES & EM
  & $1/\tau$ & ES & EM
  & $1/\tau$ & Acc
  & $1/\tau$ & Acc \\
  \midrule
  Full & 1.0 & 65.03 & 40.5 & 1.0 & 65.03 & 40.5 & 1.0 & 49.11 & 1.0 & 49.11 \\
  No Context & $\infty$ & 43.85 & 14.0 & $\infty$ & 43.85 & 14.0 & $\infty$ & 37.50 & $\infty$ & 37.50 \\
  \midrule
  Selective-Context & 3.27 & 53.68 & 24.5 & 7.49 & 49.89 & 17.5 & 2.71 & 51.33 & 6.69 & 46.43 \\
  LLMLingua2 & 3.95 & 48.09 & 15.0 & 7.89 & 45.53 & 13.5 & 3.70 & 43.36 & 5.46 & 39.82  \\
  RAG & 3.32 & 58.21 & 31.0 & 6.78 & 56.97 & 28.5 & 3.44 & 53.98 & 6.65 & 53.57 \\
  LongCodeZip & 3.96 & 56.72 & 25.5 & 6.53 & 54.91 & 23.0 & 3.29 & 51.33 & 7.49 & 50.91 \\
  \rowcolor{blue!10}
  \textbf{\approach} & \textbf{4.22} & \textbf{57.71} & \textbf{31.0} & \textbf{8.13} & \textbf{56.73} & \textbf{28.5} & \textbf{9.98} & \textbf{56.25} & \textbf{14.68} & \textbf{55.75} \\
  \bottomrule
  \end{tabular}
  \caption{Results on Long Code Completion and Long Code QA with Seed-Coder-8B-Instruct. \approach demonstrates consistent effectiveness across different model architectures, achieving superior compression while maintaining competitive task performance.}
  \label{tab:seedcoder-results}
\end{table*}

The results with Seed-Coder-8B-Instruct exhibit similar patterns to those observed with Qwen2.5-Coder-7B-Instruct. On Long Code Completion, \approach achieves 8.13x compression under 8x constraint with 56.73 ES and 28.5 EM, outperforming Selective-Context (7.49x/49.89 ES/17.5 EM) and LongCodeZip (6.53x/54.91 ES/23.0 EM). On Long Code QA, the advantages are particularly pronounced: under 8x constraint, our method achieves 14.68x compression with 55.75\% accuracy, substantially exceeding all baselines including RAG (6.65x/53.57\%) and LongCodeZip (7.49x/50.91\%). This cross-model consistency validates that the effectiveness of task-aware, line-level pruning is not specific to a particular model architecture but represents a generalizable approach to context compression for code understanding tasks.

\section{Syntactic Structure Preservation Analysis}
\label{app:ast_analysis}

Line-level pruning preserves syntactic structure better than token-level compression. Our pruning strategy retains semantically relevant lines with minimal syntactic context for code integrity, while token-level methods disrupt AST structure through arbitrary token deletion. We evaluate AST correctness using tree-sitter~\citep{treesitter} on Long Code Completion. \Cref{tab:ast_correctness} presents results.

\begin{table}[t]
\centering
\label{tab:ast_correctness}
% \resizebox{\columnwidth}{!}{
    \begin{tabular}{lc}
    \toprule
    \textbf{Method} & \textbf{AST Correctness (\%)} \\
    \midrule
    \multicolumn{2}{l}{\textit{Baseline (No Compression)}} \\
    Full Context & 98.5 \\
    No Context & 98.5 \\
    \midrule
    \multicolumn{2}{l}{\textit{Token-Level Compression}} \\
    LLMLingua2 & 0.29 \\
    Selective Context & 12.4 \\
    % Function RAG + Random Token Pruner & 49.6 \\
    Random Token Pruner & 49.6 \\
    \midrule
    \multicolumn{2}{l}{\textit{Line-Level Compression}} \\
    Function RAG & 92.3 \\
    \quad + Random Line Pruner & 78.2 \\
    % \rowcolor{blue!10}
    \quad + \textbf{\approach} & 87.3 \\
    \midrule
    \multicolumn{2}{l}{\textit{Structural Compression}} \\
    LongCodeZip & 89.3 \\
    \quad + \textbf{\approach} & 76.8 \\
    % \midrule
    % \multicolumn{2}{l}{\textit{Ablation}} \\
    \bottomrule
    \end{tabular}
% }
\caption{AST Correctness Rate after Compression. Line-level methods maintain substantially higher syntactic validity compared to token-level approaches.}
\end{table}

Token-level methods (LLMLingua2, Selective Context) achieve near-zero AST correctness (0.29\%, 12.4\%), while line-level approaches maintain substantially higher validity. Our \approach achieves 87.3\% AST correctness on Function RAG, outperforming random token pruning (49.6\%) and competitive with random line pruning (78.2\%). This demonstrates that structure-aware labeling successfully identifies safe removals while preserving syntactic dependencies. Note that \approach operates as a second-stage pruner on top of Function RAG's output, meaning any additional compression inherently introduces some risk of removing syntactically critical lines. The slight reduction from baseline Function RAG (92.3\%) to our method (87.3\%) reflects this compression-validity trade-off, where the 5\% decrease enables substantially higher compression ratios while maintaining practical code validity for downstream tasks.

\section{Case Study on SWE Bench}
\label{app:case_study}

To complement the aggregate statistics from SWE-Bench experiments, we conduct an in-depth case study comparing agent behaviors under two configurations: a Baseline agent operating with full interaction histories, and a Pruner-augmented agent that applies context pruning to file observations. Both agents operate on the same set of software engineering tasks from SWE-Bench, and we analyze their trajectories through the lens of tool invocation patterns and token consumption. We select two representative instances that illustrate distinct benefits of context pruning: one where pruning enables task completion by preventing resource exhaustion, and another where pruning achieves structural efficiency gains even when both agents succeed.

\subsection{Case 1: High-Impact Scenario}

The first case examines task \texttt{django\_\_django-10554}, which requires fixing a missing deep copy in the \texttt{Query.clone()} method for the \texttt{combined\_queries} attribute. As shown in~\Cref{tab:case1}, the Baseline agent exhausts its resource limits after 164 steps, accumulating over 7 million tokens with a maximum prompt length of 87,790 tokens. In contrast, the Pruner-augmented agent completes the task successfully in 56 steps with 1.17 million tokens and a peak prompt length of 38,226 tokens. The token reduction reaches 83.3\% in absolute terms, demonstrating that context pruning can transform resource-bound failures into successful completions.

\begin{table}[htbp]
\centering
\label{tab:case1}
% \resizebox{\columnwidth}{!}{
\begin{tabular}{lrrrrrrr}
\toprule
\textbf{Setting} & \textbf{Steps} & \textbf{Read} & \textbf{Search} & \textbf{Exec} & \textbf{Edit} & \textbf{Tokens} \\
\midrule
Baseline & 164 & 59 & 39 & 1 & 25 & 7,001,934 \\
Pruner & 56 & 20 & 10 & 0 & 11 & 1,170,160 \\
\bottomrule
\end{tabular}
\caption{Comparison of agent behaviors on \texttt{django\_\_django-10554}. Baseline exhausts limits while Pruner succeeds with 83.3\% token reduction.}
% }
\end{table}

Examining the trajectory reveals fundamental differences in exploration strategy. The Baseline agent engages in extensive breadth-first file reading, issuing commands such as \texttt{find . -name "*.py" | grep union} followed by repeated segmented reads using \texttt{sed -n 'x,y'} across numerous files. This results in accumulation of redundant context from tangentially related code. Meanwhile, the Pruner agent directly navigates to the core file \texttt{django/db/models/sql/query.py}, reads it with line numbers (\texttt{cat -n}), and identifies the relevant branch (\texttt{if self.combinator: ...}) for modification. By filtering out irrelevant sections during file reads, the pruner enables the agent to maintain a focused working context, avoiding the context overflow that derails the Baseline.

\subsection{Case 2: Structural Efficiency Gains}

The second case analyzes task \texttt{django\_\_django-11740}, which requires adding foreign key dependency tracking to the migration autodetector's \texttt{AlterField} operation. Unlike the previous case, both agents successfully complete this task, but the Pruner achieves substantial efficiency improvements as shown in~\Cref{tab:case2}. Although the Pruner agent takes 6 additional steps (48 versus 42), its token consumption is 6.0\% lower and its maximum prompt length is reduced by 30.2\%. This illustrates that pruning benefits extend beyond preventing failures to improving resource efficiency in successful trajectories.

\begin{table}[htbp]
\centering
\label{tab:case2}
% \resizebox{\columnwidth}{!}{
\begin{tabular}{lrrrrrrr}
\toprule
\textbf{Setting} & \textbf{Steps} & \textbf{Read} & \textbf{Search} & \textbf{Exec} & \textbf{Edit} & \textbf{Tokens} \\
\midrule
Baseline & 42 & 12 & 8 & 1 & 18 & 857,371 \\
Pruner & 48 & 14 & 8 & 0 & 13 & 806,220 \\
\bottomrule
\end{tabular}
\caption{Comparison on \texttt{django\_\_django-11740}. Both succeed, but Pruner achieves 30.2\% reduction in peak prompt length.}
% }
\end{table}

Analysis of the trajectories shows that both agents converge on the same logical solution, modifying \texttt{autodetector.py} to adjust the \texttt{\_get\_dependencies\_for\_foreign\_key} invocation. However, their reading strategies differ qualitatively. The Baseline agent performs multiple segmented reads of the target file and creates temporary validation scripts (e.g., \texttt{python /tmp/test\_uuid\_to\_fk3.py}) to verify understanding, accumulating historical noise from these exploratory steps. The Pruner agent reads the complete file with line numbers once, directly edits the relevant section, and avoids auxiliary validation artifacts. This behavioral shift reflects a transition from defensive exploration to confident intervention, enabled by cleaner, more focused context at each decision point.

\paragraph{Discussion}
These cases illustrate two distinct modes of benefit from context pruning. In the high-impact scenario, pruning prevents catastrophic context overflow that leads to task abandonment, effectively converting failures into successes. In the structural efficiency scenario, pruning does not change task outcomes but significantly reduces the computational burden and maximum context requirements, improving operational cost-effectiveness. Both patterns support the hypothesis that context pruning serves not merely as an engineering optimization but as a cognitive strategy that enables more efficient problem-solving behaviors. By reducing the volume of irrelevant information presented to the agent at each step, pruning allows the underlying language model to allocate more attention to task-critical signals, thereby improving both decision quality and resource utilization.

\section{Prompt Templates}
\label{app:prompts}

This appendix presents the complete prompt templates used in our experiments. We provide four key prompt templates that form the core of our approach. First, the Silver Label Prompt is used to generate training data by asking models to answer queries using provided code context with explicit line citations. Second, the Mini SWE Agent with Pruner Template defines the agent system for solving software engineering tasks, which includes detailed instructions on response format, context focus questions, workflow recommendations, and command execution rules. Third, the SWE-QA Bash Tool Descriptions specify how the bash execution tool works within the agent framework, particularly highlighting the usage of context focus questions for filtering large outputs. Finally, the Quality Evaluation Prompt provides the criteria and procedure for assessing the quality of pruned code samples across three dimensions: query quality, deletion relevance, and semantic preservation.

\clearpage

\begin{figure*}[t!]
\begin{tcolorbox}[title=Silver Label Prompt]
\begin{lstlisting}[basicstyle=\scriptsize\ttfamily, breaklines=true, columns=fullflexible]
You are given:
- a natural-language(or code for code completion task) Query
- a code snippet split into numbered lines (1>, 2>, 3>, ...)

Question: {query}
Code Context:
{code}

Answer the Question, using ONLY information provided in the Code Context. If no useful information
is provided, you MUST output "No answer". If some parts of the Context are used to answer, you
MUST cite ALL the corresponding lines. 

Use the symbols [ ] to indicate when a fact comes from a line in the context, e.g [1] for a fact from line 1. 
- For multi-line context, use [line1-line2], e.g. [12-25]. 
- For multi context, use [line1,line2,...], e.g. [1,3,5-7].

You should only answer the given question and should not provide any additional information

HINT: 
- For code, context should be wider than `the line just answer the question`, for example, if the question is about a variable in a class method function, include the function definition, class definition and where it is used.
- When you try to cite something, its better to cite the structure of the code. 
e.g. if you want to cite B1 in the code structure below:
```
1> if cond:
2>    A1
3>    A2
4> else:
5>    B1
```
, best citation will be [1,4,5], which keeps the structure of the `if-else` block while removing the unrelated A1, A2.

Now give your answer with citations:
\end{lstlisting}
\end{tcolorbox}
\end{figure*}

\FloatBarrier
\clearpage
    
\begin{figure*}[t!]
\begin{tcolorbox}[title=Mini SWE Agent with pruner template]
\begin{lstlisting}[basicstyle=\scriptsize\ttfamily, breaklines=true, columns=fullflexible]
agent:
  system_template: |
    You are a helpful assistant that can interact multiple times with a computer shell to solve programming tasks.

    ## Response Format

    Your response must contain exactly ONE bash code block(with optinal context_focus_question shown below) with ONE command (or commands connected with && or ||).
    Include a THOUGHT section before your command where you explain your reasoning process.

    <format_example>
    THOUGHT: Your reasoning and analysis here. Explain why you want to perform the action.

    ```bash
    your_command_here
    ```
    <context_focus_question>
    Optional question to focus on relevant parts of the command output.
    </context_focus_question>
    </format_example>

    ## Context Focus Question (Optional)

    The `context_focus_question` is an optional field that helps filter large command outputs to show only relevant information.

    **Question requirements:**
    - question **MUST be a complete, self-contained question** (not keywords or phrases)
    - question should be specific enough to filter effectively
    - questions should **NOT** contain file-level info (e.g., filenames, line numbers), as the filter model only processes direct command output. For such info, use tools like grep/sed and leave the question blank.
    - question should be place it immediately after the bash code block (after the closing ```)

    ### Examples of a good context_focus_question:
    - Find where the [some logic] is implemented in the [some class/function]?
    - Given [some background], [some problem]?
    - Locate the [some logic]?
    - How can the code implement [some feature]?
    - [some combination of the above, like background + origin problem + current purpose + specific attention area]
    
    ### Examples of a bad context_focus_question:
    - load_raw function (too vague)
    - lines 50-100 of data_loader.py (contains file-level info)
    - fix the bug in rwkv6.py (too vague)

    Gives some context for the query is better for more effective filtering.
    Failure to follow these rules will cause your response to be rejected.
  instance_template: |
    <pr_description>
    Consider the following PR description:
    {{task}}
    </pr_description>

    <instructions>
    # Task Instructions

    ## Overview
    You're a software engineer interacting continuously with a computer by submitting commands.
    You'll be helping implement necessary changes to meet requirements in the PR description.
    Your task is specifically to make changes to non-test files in the current directory in order to fix the issue described in the PR description in a way that is general and consistent with the codebase.

    IMPORTANT: This is an interactive process where you will think and issue ONE command, see its result, then think and issue your next command.

    For each response:
    1. Include a THOUGHT section explaining your reasoning and what you're trying to accomplish
    2. Provide exactly ONE bash command to execute
    3. Give an optional `context_focus_question` to filter the command output(or not give for output no need to be filtered)

    ## Important Boundaries
    - MODIFY: Regular source code files in /testbed (this is the working directory for all your subsequent commands)
    - DO NOT MODIFY: Tests, configuration files (pyproject.toml, setup.cfg, etc.)

\end{lstlisting}
\end{tcolorbox}
\end{figure*}
    
\clearpage

\begin{figure*}[p]
\centering
\begin{tcolorbox}[title=Mini SWE Agent with pruner template (Part 2)]
\begin{lstlisting}[basicstyle=\scriptsize\ttfamily, breaklines=true, columns=fullflexible]
    ## Recommended Workflow
    1. Analyze the codebase by finding and reading relevant files to figure out why this issue happenes(with context_focus_question for large outputs like read a whole file)
    **HINT**: since we have pruner enabled, prefer reading files fully instead of using grep/find first
    **ATTENTION**: pruned code might miss some details, so its **highly recommend** to use `cat -n` or `nl -ba` command with context_focus_question, so you can see the line numbers. If you looked the filtered output and need more context, you can always use `sed` without context_focus_question to read more lines before/after the relevant lines.

    2. When you have enough information, edit the source code to resolve the issue, you have only one chance, no tests, no retry, so read widely first(with context_focus_questions) before making changes.

    ## Command Execution Rules
    You are operating in an environment where
    1. You write a single command
    2. The system executes that command in a subshell
    3. You see the result(if you set a context_focus_question, the output will be filtered accordingly)
    4. You write your next command

    Each response should include:
    1. A **THOUGHT** section where you explain your reasoning and plan
    2. A single bash code block with your command
    3. An optional `context_focus_question` to filter the command output, actually you can put some thoughts directly as background in it.

    Format your responses like this:

    <format_example>
    THOUGHT: Here I explain my reasoning process, analysis of the current situation,
    and what I'm trying to accomplish with the command below.

    ```bash
    your_command_here
    ```
    <context_focus_question>
    Your question here.
    </context_focus_question>
    </format_example>

    Commands must be specified in a single bash code block:

    ```bash
    your_command_here
    ```

    **CRITICAL REQUIREMENTS:**
    - Your response SHOULD include a THOUGHT section explaining your reasoning
    - Your response MUST include EXACTLY ONE bash code block(with optinal context_focus_question)
    - This bash block MUST contain EXACTLY ONE command (or a set of commands connected with && or ||)
    - If you include zero or multiple bash blocks, or no command at all, YOUR RESPONSE WILL FAIL
    - Do NOT try to run multiple independent commands in separate blocks in one response
    - Directory or environment variable changes are not persistent. Every action is executed in a new subshell.
    - However, you can prefix any action with `MY_ENV_VAR=MY_VALUE cd /path/to/working/dir && ...` or write/load environment variables from files

    Example of a CORRECT response(no context_focus_question because `ls` will not output large content, and we want to see the full output):
    <example_response>
    THOUGHT: I need to understand the structure of the repository first. Let me check what files are in the current directory to get a better understanding of the codebase.

    ```bash
    ls -la
    ```
    </example_response>

    Example of an INCORRECT response:
    <example_response>
    THOUGHT: I need to examine the codebase and then look at a specific file. I'll run multiple commands to do this.

    ```bash
    ls -la
    ```


  \end{lstlisting}
\end{tcolorbox}
\end{figure*}

\begin{figure*}[p]
\centering
\begin{tcolorbox}[title=Mini SWE Agent with pruner template (Part 3)]
\begin{lstlisting}[basicstyle=\scriptsize\ttfamily, breaklines=true, columns=fullflexible]
    Now I'll read the file:

    ```bash
    cat -n file.txt
    ```
    <context_focus_question>
    Some question about your purpose for reading the file.
    </context_focus_question>

    </example_response>

    If you need to run multiple commands, either:
    1. Combine them in one block using && or ||
    ```bash
    command1 && command2 || echo "Error occurred"
    ```  

    2. Wait for the first command to complete, see its output, then issue the next command in your following response.

    ## Environment Details
    - You have a full Linux shell environment
    - Always use non-interactive flags (-y, -f) for commands
    - Avoid interactive tools like vi, nano, or any that require user input
    - If a command isn't available, you can install it

    ## Useful Command Examples

    ### Create a new file:
    ```bash
    cat <<'EOF' > newfile.py
    import numpy as np
    hello = "world"
    print(hello)
    EOF
    ```

    ### Edit files with sed:
    ```bash
    # Replace all occurrences
    sed -i 's/old_string/new_string/g' filename.py

    # Replace only first occurrence
    sed -i 's/old_string/new_string/' filename.py

    # Replace first occurrence on line 1
    sed -i '1s/old_string/new_string/' filename.py

    # Replace all occurrences in lines 1-10
    sed -i '1,10s/old_string/new_string/g' filename.py
    ```

    ### View file content:
    ```bash
    # View specific lines with numbers
    nl -ba filename.py | sed -n '10,20p'
    <context_focus_question>
    If you know what line you want to view and the range of `sed` if large, you can also set a question here. 
    </context_focus_question>
    ```

    ### Any other command you want to run
    ```bash
    anything
    ```

    ## Submission
    When you've completed your work (reading, editing, testing), and cannot make further progress
    issue exactly the following command:

    ```bash
    echo COMPLETE_TASK_AND_SUBMIT_FINAL_OUTPUT && git add -A && git diff --cached
    ```

    
  \end{lstlisting}
\end{tcolorbox}
\end{figure*}

\clearpage

\begin{figure*}[p]
\centering
\begin{tcolorbox}[title=Mini SWE Agent with pruner template (Part 4)]
\begin{lstlisting}[basicstyle=\scriptsize\ttfamily, breaklines=true, columns=fullflexible]
    This command will submit your work.
    You cannot continue working (reading, editing, testing) in any way on this task after submitting.
    
    </instructions>
  action_observation_template: |
    <returncode>{{output.returncode}}</returncode>
    {% if output.output | length < 10000 -%}
    <output>
    {{ output.output -}}
    </output>
    {%- else -%}
    <warning>
    The output of your last command was too long.
    Please try a different command that produces less output.
    If you have not used context_focus_question, use it.
    If you have used it but the output is still too long, do not try to change the question, just follow the instructions below while also use the context_focus_question.
    If you're looking at a file you can try use head, tail or sed to view a smaller number of lines selectively.
    If you're using grep or find and it produced too much output, you can use a more selective search pattern.
    If you really need to see something from the full command's output, you can redirect output to a file and then search in that file.
    </warning>
    {%- set elided_chars = output.output | length - 10000 -%}
    <output_head>
    {{ output.output[:5000] }}
    </output_head>

  
    <elided_chars>
    {{ elided_chars }} characters elided
    </elided_chars>
    <output_tail>
    {{ output.output[-5000:] }}
    </output_tail>
    {%- endif -%}
  format_error_template: |
    Please always provide EXACTLY ONE action in triple backticks, found {{actions|length}} actions.

    Please format your action in triple backticks as shown in <response_example>.

    <response_example>
    THOUGHT: Here are some thoughts about why you want to perform the action.

    ```bash
    <action>
    ```
    <context_focus_question>
    Please provide a question about the context you want to focus on. This part is optional. If you don't need to filter the output, you can leave it blank or omit it.
    </context_focus_question>
    </response_example>

    If you have completed your assignment, please consult the first message about how to
    submit your solution (you will not be able to continue working on this task after that).

    Note: A common trigger for this error is you put > 1 bash block(action) in your response.
  step_limit: 250
  cost_limit: 3.
  pruner:
    url: http://localhost:8000/prune
    timeout: 120
    retries: 3
    min_chars: 500
    chunk_overlap_tokens: 50
    threshold: 0.5

environment:
  cwd: "/testbed"
  timeout: 60
  env:
    PAGER: cat
    MANPAGER: cat
    LESS: -R
    PIP_PROGRESS_BAR: 'off'
    TQDM_DISABLE: '1'
  environment_class: docker

\end{lstlisting}
\end{tcolorbox}
\end{figure*}

\clearpage
\FloatBarrier

\begin{figure*}[t!]
\begin{tcolorbox}[title=SWE-QA bash tool descriptions]
\begin{lstlisting}[basicstyle=\scriptsize\ttfamily, breaklines=true, columns=fullflexible]
_BASH_DESCRIPTION = """Execute bash commands in the workspace.
* Use this tool to run shell commands, navigate directories, search files, etc.
* Commands are executed in the workspace directory.
* Output is returned as text.
"""

_BASH_DESCRIPTION_PRUNE = (
    _BASH_DESCRIPTION
    + """
The meaning of argument `context_focus_question` (Optional):

Use `context_focus_question` to filter large command outputs for relevant information.

**Requirements:**
- Must be a complete, self-contained question (not keywords/phrases)
- Be specific for effective filtering
- Don't include file-level info (filenames, line numbers) - use grep/sed instead

**Good examples:**
- Where is [some logic] implemented in [some class/function]?
- Given [background], what's the [problem]?
- How does the code implement [feature]?

**Bad examples:**
- load_raw function (too vague)
- lines 50-100 of data_loader.py (contains file info)
- fix bug in rwkv6.py (too vague)

**IMPORTANT:** With pruner enabled, prefer `cat -n` or `nl -ba` with context_focus_question to see line numbers. Then you can use `sed` without filtering for more detailed context since you have line number information.

**IMPORTANT:** If the command output is small and important like `ls`, just leave context_focus_question blank.
"""
)
\end{lstlisting}
\end{tcolorbox}
\end{figure*}
  
\begin{figure*}[t!]
\begin{tcolorbox}[title=Quality Evaluation Prompt]
\begin{lstlisting}[basicstyle=\scriptsize\ttfamily, breaklines=true, columns=fullflexible]
You are a code quality evaluator for a code pruning dataset. Your task is to assess the quality of a query-guided code deletion sample.

You will be given:
1. **Query**: A code snippet for means making completion below or a natural language question/task
2. **Original Code**: Full code snippet with line numbers
3. **Diff**: Shows which lines were removed (- prefix) and kept (no change or + prefix)

Evaluate THREE dimensions:

## 1. Query Quality
- **Good**: Realistic, specific, actionable developer question related to a partial feature/function
- **Acceptable**: Valid but generic, or slightly unclear but answerable
- **Poor**: Too vague, treats code as the subject ("explain this code")

HINT: just focus on query itself, don't take query-code relevance into consideration in this part.

## 2. Deletion Relevance
- **Appropriate**: Removes truly unrelated code while keeping necessary context
- **Minimal**: Mostly removes whitespace/comments/trivial lines, little semantic pruning
- **Excessive**: Removes too much, including code relevant to the query

HINT: Both query-code high relevance and low relevance are ok, key point is the context preserved correctly. For high relevance, might more code; For low relevance, might less code.

## 3. Semantic Preservation
- **Preserved**: Remaining code is syntactically valid and semantically coherent (can understand the query-relevant logic)
- **Partially Preserved**: Minor issues (e.g., unmatched braces, missing imports that don't affect core logic understanding)
- **Broken**: Code is syntactically invalid or key logic is incomprehensible

## Overall Quality Rating
Based on the above three dimensions, assign:
- **high**: All three dimensions are good/appropriate/preserved, or at most one acceptable/partially_preserved
- **medium**: Two dimensions are good/acceptable/appropriate, one has issues; or all three are acceptable
- **low**: Two or more dimensions are poor/minimal/broken, or query is fundamentally flawed

---

### Input Data:

**Query:**
{query}

**Original Code (with line numbers):**
{code_with_numbers}

**Diff (deletions marked with -):**
```diff
{diff}
```

---

### Your Task:
1. Provide concise reasoning for each dimension (1 sentences per dimension)
2. Assign ratings: query_quality (good/acceptable/poor), deletion_relevance (appropriate/minimal/excessive), semantic_preservation (preserved/partially_preserved/broken)
3. Determine overall_quality (low/medium/high)

Output JSON format (no code fences, just JSON):
{{
  "reasoning": "<Brief analysis covering all three dimensions>",
  "query_quality": "<good|acceptable|poor>",
  "deletion_relevance": "<appropriate|minimal|excessive>",
  "semantic_preservation": "<preserved|partially_preserved|broken>",
  "overall_quality": "<low|medium|high>"
}}
\end{lstlisting}
\end{tcolorbox}
\end{figure*}

\end{document}